\documentstyle[12pt,epsfig]{article}

\catcode`\@=11
\def\@citex[#1]#2{%
\if@filesw \immediate \write \@auxout {\string \citation {#2}}\fi
\@tempcntb\m@ne \let\@h@ld\relax \def\@citea{}%
\@cite{%
  \@for \@citeb:=#2\do {%
    \@ifundefined {b@\@citeb}%
      {\@h@ld\@citea\@tempcntb\m@ne{\bf ?}%
      \@warning {Citation `\@citeb ' on page \thepage \space undefined}}%
      {\@tempcnta\@tempcntb \advance\@tempcnta\@ne%
      \@tempcntb\number\csname b@\@citeb \endcsname \relax%
      \ifnum\@tempcnta=\@tempcntb 
        \ifx\@h@ld\relax%
          \edef \@h@ld{\@citea\csname b@\@citeb\endcsname}%
        \else%
          \edef\@h@ld{\ifmmode{-}\else--\fi\csname b@\@citeb\endcsname}%
        \fi%
      \else
        \@h@ld\@citea\csname b@\@citeb \endcsname%
        \let\@h@ld\relax%
      \fi}%
    \def\@citea{,\penalty\@highpenalty\,}%
  }\@h@ld
}{#1}}

\def\@citeb#1#2{{[#1]\if@tempswa , #2\fi}}
\def\@citeu#1#2{{$^{#1}$\if@tempswa , #2\fi }}
\def\@citep#1#2{{#1\if@tempswa , #2\fi}}

\def\bcites{         
        \catcode`\@=11
        \let\@cite=\@citeb
        \catcode`\@=12
}

\def\upcites{         
        \catcode`\@=11
        \let\@cite=\@citeu
        \catcode`\@=12
}

\def\plaincites{      
        \catcode`\@=11
        \let\@cite=\@citep
        \catcode`\@=12
}

\newcount\hour
\newcount\minute
\newtoks\amorpm
\hour=\time\divide\hour by 60
\minute=\time{\multiply\hour by 60 \global\advance\minute by-\hour}
\edef\standardtime{{\ifnum\hour<12 \global\amorpm={am}%
        \else\global\amorpm={pm}\advance\hour by-12 \fi
        \ifnum\hour=0 \hour=12 \fi
        \number\hour:\ifnum\minute<10 0\fi\number\minute\the\amorpm}}
\edef\militarytime{\number\hour:\ifnum\minute<10 0\fi\number\minute}

\def\draftlabel#1{{\@bsphack\if@filesw {\let\thepage\relax
   \xdef\@gtempa{\write\@auxout{\string
      \newlabel{#1}{{\@currentlabel}{\thepage}}}}}\@gtempa
   \if@nobreak \ifvmode\nobreak\fi\fi\fi\@esphack}
        \gdef\@eqnlabel{#1}}
\def\@eqnlabel{}
\def\@vacuum{}
\def\marginnote#1{}
\def\draftmarginnote#1{\marginpar{\raggedright\scriptsize\tt#1}}
\def\draft{
        \pagestyle{plain}
        \overfullrule=2pt
        \oddsidemargin -.5truein
        \def\@oddhead{\sl \phantom{\today\quad\militarytime} \hfil
        \smash{\Large\sl DRAFT} \hfil \today\quad\militarytime}
        \let\@evenhead\@oddhead
        \let\label=\draftlabel
        \let\marginnote=\draftmarginnote
        \def\ps@empty{\let\@mkboth\@gobbletwo
        \def\@oddfoot{\hfil \smash{\Large\sl DRAFT} \hfil}
        \let\@evenfoot\@oddhead}
        \def\@eqnnum{(\theequation)\rlap{\kern\marginparsep\tt\@eqnlabel}%
        \global\let\@eqnlabel\@vacuum}  }

\def\blackfonts{
        \font\blackboard=msbm10 scaled\magstep1
        \font\blackboards=msbm8
        \font\blackboardss=msbm6
}

\def\nblack{            
        \def\ZZ{{Z \n{10} Z}}
        \def\NN{{N \n{14} N}}
        \def\CC{{C \n{11} C}}
        \def\RR{{R \n{11} R}}
        \def\QQ{{Q \n{12} Q}}
        \def\PP{{P \n{11} P}}
}

\def\prep{         
        \catcode`\@=11
        \input art10.sty
        \catcode`\@=12
        
        \let\small\null
        \def\blackfonts{
                \font\blackboard=msbm10
                \font\blackboards=msbm7
                \font\blackboardss=msbm5
        }
        \let\sl\it
        \twocolumn
        \sloppy
        \voffset=-2.54truecm
        \hoffset=-2.54truecm
        \flushbottom
        \parindent 1em
        \leftmargini 2em
        \leftmarginv .5em
        \leftmarginvi .5em
        \marginparwidth 48pt
        \marginparsep 10pt
        \setlength{\columnsep}{2truecm}
        \setlength{\textwidth}{25.4truecm}
        \setlength{\textheight}{17truecm}
        \baselineskip=16pt
        \oddsidemargin .18truein
        \evensidemargin .17truein
}

\def\eqalign#1{\null\,\vcenter{\openup\jot\m@th
  \ialign{\strut\hfil$\displaystyle{##}$&$\displaystyle{{}##}$\hfil
      \crcr#1\crcr}}\,}
\def\eqalignno#1{\displ@y \tabskip\centering
  \halign to\displaywidth{\hfil$\@lign\displaystyle{##}$\tabskip\z@skip
    &$\@lign\displaystyle{{}##}$\hfil\tabskip\centering
    &\llap{$\@lign##$}\tabskip\z@skip\crcr
    #1\crcr}}

\def\section{\@startsection {section}{1}{\z@}{3.ex plus 1ex minus
 .2ex}{2.ex plus .2ex}{\large\bf}}
\def\subsection{\@startsection{subsection}{2}{\z@}{2.75ex plus 1ex minus
 .2ex}{1.5ex plus .2ex}{\bf}}        

\def\appendix{{\newpage\section*{Appendix}}\let\appendix\section%
        {\setcounter{section}{0}
        \gdef\thesection{\Alph{section}}}\section}
\def\thefootnote{\arabic{footnote}}
\def\abstract{\if@twocolumn
\section*{Abstract}
\else 
\begin{center}
{\bf Abstract\vspace{-.5em}\vspace{0pt}}
\end{center}
\quotation
\fi}

\catcode`\@=12

\newcommand{\beq}{\begin{equation}}
\newcommand{\eeq}{\end{equation}}
\newcommand{\beqa}{\begin{eqnarray}}
\newcommand{\eeqa}{\end{eqnarray}}

\newcommand{\Z}{{\bf Z}}

\newcommand{\R}{{\bf R}}
\newcommand{\C}{{\bf C}}
\newcommand{\e}{{\rm e}}

\newcommand{\dd}{{\rm d}}
\newcommand{\elst}{{\ell_{\it st}}}
\newcommand{\elel}{{\ell_{11}}}
\newcommand{\gst}{{g_{\it st}}}
\newcommand{\MT}{{$M$ theory}~}

\def\noj#1,#2,{{\bf #1} (19#2)\ }
\def\jou#1,#2,#3,{{\sl #1\/ }{\bf #2} (19#3)\ }
\def\ann#1,#2,{{\sl Ann.\ Physics\/ }{\bf #1} (19#2)\ }
\def\cmp#1,#2,{{\sl Comm.\ Math.\ Phys.\/ }{\bf #1} (19#2)\ }
\def\ma#1,#2,{{\sl Math.\ Ann.\/ }{\bf #1} (19#2)\ }
\def\ng#1,#2,{{\sl Nagoya.\ Math.\ J.\/ }{\bf #1} (19#2)\ }
\def\jd#1,#2,{{\sl J.\ Diff.\ Geom.\/ }{\bf #1} (19#2)\ }
\def\invm#1,#2,{{\sl Invent.\ Math.\/ }{\bf #1} (19#2)\ }
\def\cq#1,#2,{{\sl Class.\ Quantum Grav.\/ }{\bf #1} (19#2)\ }
\def\cqg#1,#2,{{\sl Class.\ Quantum Grav.\/ }{\bf #1} (19#2)\ }
\def\ijmp#1,#2,{{\sl Int.\ J.\ Mod.\ Phys.\/ }{\bf A#1} (19#2)\ }
\def\jmphy#1,#2,{{\sl J.\ Geom.\ Phys.\/ }{\bf #1} (19#2)\ }
\def\jams#1,#2,{{\sl J.\ Amer.\ Math.\ Soc.\/ }{\bf #1} (19#2)\ }
\def\grg#1,#2,{{\sl Gen.\ Rel.\ Grav.\/ }{\bf #1} (19#2)\ }
\def\mpl#1,#2,{{\sl Mod.\ Phys.\ Lett.\/ }{\bf A#1} (19#2)\ }
\def\nc#1,#2,{{\sl Nuovo Cim.\/ }{\bf #1} (19#2)\ }
\def\np#1,#2,{{\sl Nucl.\ Phys.\/ }{\bf B#1} (19#2)\ }
\def\pl#1,#2,{{\sl Phys.\ Lett.\/ }{\bf #1B} (19#2)\ }
\def\pla#1,#2,{{\sl Phys.\ Lett.\/ }{\bf #1A} (19#2)\ }
\def\pr#1,#2,{{\sl Phys.\ Rev.\/ }{\bf #1} (19#2)\ }
\def\prd#1,#2,{{\sl Phys.\ Rev.\/ }{\bf D#1} (19#2)\ }
\def\prl#1,#2,{{\sl Phys.\ Rev.\ Lett.\/ }{\bf #1} (19#2)\ }
\def\prp#1,#2,{{\sl Phys.\ Rept.\/ }{\bf #1C} (19#2)\ }
\def\ptp#1,#2,{{\sl Prog.\ Theor.\ Phys.\/ }{\bf #1} (19#2)\ }
\def\ptpsup#1,#2,{{\sl Prog.\ Theor.\ Phys.\/ Suppl.\/ }{\bf #1} (19#2)\ }
\def\rmp#1,#2,{{\sl Rev.\ Mod.\ Phys.\/ }{\bf #1} (19#2)\ }
\def\yadfiz#1,#2,#3[#4,#5]{{\sl Yad.\ Fiz.\/ }{\bf #1} (19#2) #3%
\ [{\sl Sov.\ J.\ Nucl.\ Phys.\/ }{\bf #4} (19#2) #5]}
\def\zh#1,#2,#3[#4,#5]{{\sl Zh.\ Exp.\ Theor.\ Fiz.\/ }{\bf #1} (19#2) #3%
\ [{\sl Sov.\ Phys.\ JETP\/ }{\bf #4} (19#2) #5]}

\def\beq{\begin{equation}}
\def\eeq{\end{equation}}
\def\beqar{\begin{eqnarray}}
\def\eeqar{\end{eqnarray}}
\def\non{\nonumber}

\newcommand{\be}{\begin{equation}}
\newcommand{\ee}{\end{equation}}
\newcommand{\bea}{\begin{eqnarray}}
\newcommand{\eea}{\end{eqnarray}}

\def\nfrac#1#2{{\displaystyle{\vphantom1\smash{\lower.5ex\hbox{\small$#1$}}%
        \over\vphantom1\smash{\raise.25ex\hbox{\small$#2$}}}}}

\def\n#1{\mskip-#1mu}

\def\to{\rightarrow}

\def\lae{\mathrel{\mathop{\smash{\lower .5 ex \hbox{$\stackrel<\sim$}}}}}
\def\lae{\mathrel{\mathop{\smash{\lower .5 ex \hbox{$\stackrel>\sim$}}}}}

\def\Tr{{\rm Tr}}
\def\l:{\mathopen{:}\,}
\def\r:{\,\mathclose{:}}

\catcode`\@=11
\def\theequation{\arabic{equation}}
\catcode`\@=12

\nblack
\bcites

\nblack

\catcode`\@=11
\def\theequation{\thesection.\arabic{equation}}
\@addtoreset{equation}{section}
\@addtoreset{footnote}{section}
\@addtoreset{footnote}{subsection}
\catcode`\@=12

\newcommand{\beqn}{\begin{equation}}
\newcommand{\eeqn}{\end{equation}}
\newcommand{\beqnarray}{\begin{eqnarray}}
\newcommand{\eeqnarray}{\end{eqnarray}}
\newcommand {\bear} [1] {\begin {array} {#1}}
\newcommand {\ear} {\end {array}}

\newcommand{\CP}{{\bf C}{\rm P}}

\newcommand {\beqarn} {\begin{eqnarray*}}
\newcommand {\eeqarn} {\end{eqnarray*}}

\begin{document}

\begin{titlepage}

\begin{center}
\today
\hfill LBNL-41198, UCB-PTH-97/68\\
\hfill                  hep-th/9801060

\vskip 1.5 cm
{\large \bf Branes and Dynamical Supersymmetry Breaking}
\vskip 1 cm 
{Jan de Boer, Kentaro Hori, Hirosi Ooguri and Yaron Oz}\\
\vskip 0.5cm
{\sl Department of Physics,
University of California at Berkeley\\
366 Le\thinspace Conte Hall, Berkeley, CA 94720-7300, U.S.A.\\
and\\
Theoretical Physics Group, Mail Stop 50A--5101\\
Ernest Orlando Lawrence Berkeley National Laboratory\\
Berkeley, CA 94720, U.S.A.\\}

\end{center}

\vskip 0.5 cm

\begin{abstract}
We study dynamical supersymmetry breaking in four dimensions 
using the fivebrane of $M$ theory, in particular for the
Izawa-Yanagida-Intriligator-Thomas (IYIT) model, which we 
realize as the worldvolume theory of a
certain M-theory fivebrane configuration. 
 From the brane point of view, supersymmetry is 
broken when a holomorphic configuration with the proper boundary conditions does not exist.
We discuss the difference 
between explicit and spontaneous supersymmetry breaking
and between runaway behavior and having a stable vacuum.
As a preparation for the study of the IYIT model,
we examine a realization of the orientifold four-plane
in $M$ theory. We derive known as well as new results on the moduli 
spaces of $N=2$ and $N=1$ theories with symplectic
gauge groups. These results are based on a hypothesis that a certain
intersection of the fivebrane and the $\Z_2$ fixed plane
breaks supersymmetry.
In the IYIT model, we show that the brane exhibits runaway 
behavior when the flavor group is gauged.
On the other hand, if the flavor group is not gauged,
we find that the brane does not run away. 
We suggest that a stable supersymmetry-breaking vacuum
is realized in the region beyond the reach of the
supergravity approximation.

\end{abstract}

\end{titlepage}

\section{Introduction}

One of the most important and interesting issues in supersymmetric
gauge theories is the dynamical breaking of supersymmetry.
Under what conditions and by what mechanisms
is supersymmetry dynamically broken?
If supersymmetry is broken, is there a non-supersymmetric
stable vacuum? If there is, what is the vacuum energy?
There have been a lot of important works from various point of view
concerning such questions
but they still remain as difficult and fascinating as before.

If there is a new method for analyzing field theory, it is worth
examining whether it
sheds a new light on the issue of dynamical supersymmetry breaking.
The study of the worldvolume dynamics of branes in string theory and \MT
has given a new perspective on the study of the
strong coupling dynamics of supersymmetric gauge theories
in various dimensions
\cite{KLMVW,HW,g1,n1,Barbon,witten1,ejs,n2,n3,n4,g2,n5,n6,n7,lll,BSTY,n8,HOO,W,n9,tau,n10,n11,n12,n13,Yi,Nakatsu,SS,n14,do,LL,Mikh,CS,n15,AOT,gipe,LPT,AOT1,volo,BarP,Fayy,HLW,kah,tera,n16,n17,ggkk,n18}.
The worldvolume theory
depends on parameters such as the string coupling constant
which are absent in the ordinary gauge theory.
 For generic values of such parameters, it is different from the
gauge theory since it interacts with the bulk degrees of freedom
in the ten or eleven-dimensional space-time. In addition,
it contains Kaluza-Klein modes associated with the compactification
of the worldvolume theory to lower dimensions.
However, there are certain quantities which are independent of
such differences, and one can obtain exact results by
going to the region of the parameter space where the worldvolume
dynamics simplifies, e.g., the eleven-dimensional supergravity limit of
$M$ theory. Examples of such quantities are holomorphic
or BPS objects such as
the effective holomorphic gauge coupling constant \cite{witten1,lll,BSTY},
vevs of some chiral operators \cite{HOO,W,tau,do,gipe}, and
the mass or tension of BPS states \cite{W,Yi,Mikh,Fayy,volo}.
 Furthermore, when the universality class
is expected to remain the same, we can make  predictions
about qualitative features of the theory
which might be harder to obtain  
from conventional field theory methods \cite{W}.
Therefore, it is interesting to see whether 
branes can provide a new point of view on dynamical
supersymmetry breaking.

In this paper, we construct \MT fivebrane
realizations of four-dimensional
gauge theories that exhibit dynamical
supersymmetry breaking. Some of them are expected to have a
stable non-supersymmetric vacuum, the others are not. 
We study how the existence or absence of a stable vacuum is
realized in the fivebrane picture.
A field theory model of dynamical supersymmetry
breaking without chiral matter and with a stable non-supersymmetric vacuum
was constructed recently by Izawa-Yanagida \cite{IY}
and Intriligator-Thomas \cite{IT}
using knowledge of the strong coupling dynamics
of supersymmetric gauge theory with symplectic gauge
group \cite{Seiberg94,IP}.
Their model (which we shall call the IYIT model)
exhibits dynamical supersymmetry breaking and can be argued
to have a stable non-supersymmetric vacuum by
computing  the one-loop correction to the scalar potential.
If we gauge the orthogonal flavor group, supersymmetry is still
broken but the same one-loop computation exhibits runaway behavior and
does not show the existence of a stable vacuum. 
Using the brane picture, we will present
evidence that such a vacuum does in fact not exist.
On the other hand, if the flavor group is not gauged,
we find that the brane does not run away. 
We suggest that a stable vacuum without supersymmetry, 
if it exists, is realized in the region beyond the reach of the
supergravity approximation. 
The situation is very similar to what happens in
the field theory analysis of dynamical supersymmetry
breaking of this type. There,
one has to make an assumption about the behavior of the K\"ahler
potential in the strong coupling region of the gauge theory in order
to prove the existence of a stable vacuum
which does break supersymmetry.

Supersymmetric gauge theories with symplectic gauge group
can be realized as the worldvolume dynamics of Type IIA branes
by introducing an orientifold. To realize the theories mentioned
above, we chose to introduce an orientifold four-plane.
Although we do not have a good understanding of the
dynamics of fivebranes in \MT in the presence of
a Type IIA orientifold four-plane,
we can show that the brane reproduces information
on the moduli space of supersymmetric vacua
by introducing a simple hypothesis concerning the intersection
of the fivebrane and the $\Z_2$ fixed plane.
The hypothesis indicates the presence of a force
(which can be either finite or infinite)
between the points of intersection,
which plays a crucial role when we discuss
the issue of existence of a stable non-supersymmetric vacuum.

The paper is organized as follows.
In section 2, we present the general idea on spontaneous
supersymmetry breaking in terms of branes.
In particular, we clarify the distinction between
spontaneous and explicit breaking
and demonstrate the concepts using  a toy model.

In section 3, we examine the simple hypothesis
mentioned above by studying the brane configurations corresponding to
$N=2$ and $N=1$ supersymmetric QCD with symplectic gauge group.
We find that the brane picture together with the hypothesis 
correctly reproduce
the known results of $N=2$ and $N=1$ field
theories and yield a new result about the total
moduli space of vacua of $N=2$ theories. 
Part of the results of this section were recently obtained in
\cite{tera,AOT}. 

In section 4, we study the IYIT model in the brane framework.
We first determine the structure of supersymmetric vacua
of the model with some perturbations using ordinary field theory 
methods. Next, we realize the perturbed system via branes, using the
electric-magnetic duality as a guide. 
Then, we consider the system with supersymmetry breaking
and discuss  the existence of a stable non-supersymmetric vacuum.
Under the plausible assumption that the potential energy between a 
fivebrane and the orientifold is negligible when the distance between
the fivebrane and the orientifold is much larger than the eleven
dimensional Planck length, we show that the brane does not run away.

In section 5, we study the IYIT model with the orthogonal flavor group
gauged. We examine the space of supersymmetric vacua for general
numbers of colors and flavors using the brane based on the basic hypothesis,
and find that the brane reproduces the correct field theory results.
 For the number of colors and flavors corresponding to the IYIT model,
we will present evidence
that there is runaway behavior.

\section{Branes and Supersymmetry Breaking}

In this section we will discuss the general aspects of supersymmetry
breaking in the brane framework, and present a simple
two dimensional  model as an illustration.

\subsection{General Idea}

Supersymmetric gauge theories can be studied in various dimensions by
realizing them as theories on the world volume of branes.
Of particular importance to us in the study of supersymmetry
breaking  will be $N=1$ supersymmetric gauge theories
in four dimensions.
These are constructed using configurations of intersecting branes in
Type IIA string theory which capture the
semiclassical features of the gauge theories. In order to study the 
quantum properties we have to lift the brane configuration to \MT
\cite{witten1}.
It is described by a fivebrane wrapping a Riemann surface
holomorphically embedded in the space-time
(or more precisely, in a Calabi-Yau three-fold
which is a part of the eleven-dimensional space-time).
The Riemann surface encodes information about the
supersymmetric ground states of the theory; For example,
the effective gauge coupling constant in an abelian Coulomb phase
\cite{witten1,lll,BSTY,kah},
chiral symmetry breaking, and gaugino or monopole condensation
in a confining phase \cite{HOO,W,tau,do},
vevs of some chiral operators such as mesons and baryons
in a Higgs phase \cite{HOO,tau,gipe,AOT1,tera,AOT},
and the mass or tension of BPS states \cite{Yi,Mikh,Fayy,volo}.

The Riemann surface is a supersymmetric cycle \cite{bbs,bsv,us}
and the fivebrane wrapping the Riemann surface is a supersymmetric
configuration, namely a BPS object. 
The amount of supersymmetry preserved by the 
fivebrane configuration depends on the details of the Riemann surface.
The signal for supersymmetry breaking is that
the fivebrane is no longer wrapping a holomorphic curve but rather
a nonholomorphic
real two dimensional surface. Such a fivebrane configuration is
not a BPS object because it breaks all supersymmetry completely.
Therefore the issue of the stability of the configuration arises.

Generically we can have a stable configuration
if the two dimensional surface is
of minimal area with respect to the space-time metric.
There are however some subtleties that will arise.
The first subtlety is associated with the fact that the
brane configuration can be a factorized surface. In such a case the
disconnected components affect each other via the gravitational
force which has to be taken into 
account when analyzing the stability of the configuration. This
effect was crucial for instance in deriving Higgs branch
metrics in \cite{kah}. 
The second subtlety is associated with the orientifold
four-plane which will be used in this paper in order
to get symplectic gauge groups. The orientifold creates
a force that acts on the fivebrane intersecting with the $\Z_2$
fixed plane. Taking such an effect into account
was actually necessary in order to obtain the hyperelliptic 
curves describing the $N=2$ Coulomb branches of
gauge theories based on
symplectic and orthogonal gauge groups \cite{lll}.
We will see in section 3
that this is also required in order for
$N=1$ fivebrane configurations 
to correctly describe the supersymmetric ground states
of the gauge field theories.
Clearly this effect should also be taken into account when
studying the stability of the
non-supersymmetric fivebrane configurations.

The holomorphic fivebrane configurations are of minimal volume
in their homology class. They correspond to zero energy vacua
of the field theory.
The non-holomorphic configurations have a larger volume.
In the absence of the forces that we
discussed above the difference between the volumes of the
non-holomorphic and holomorphic 
configurations corresponds to the non-zero energy of the
non-supersymmetric  vacuum.
When the above forces are relevant their contribution to the
potential energy  has to be taken into account.

There is a difference between explicit
and spontaneous supersymmetry breaking.
Although in both cases the two dimensional surface is not holomorphic 
there is a difference between the asymptotic boundary conditions at
infinity of the surface.

\vspace{-0.1cm}
The fivebrane we are considering is wrapped on a
two-dimensional surface which is non-compact in one direction.
In order to realize a four-dimensional gauge theory
from the six-dimensional worldvolume theory, we must
restrict the allowed motion of the brane by specifying
a set of boundary conditions at infinity.
In general, a symmetry of the original theory is also
a symmetry of the restricted system
when it preserves the boundary condition
at infinity, whereas it is explicitly or anomalously
broken when it breaks the boundary condition.
When a symmetry remains, it is unbroken by a choice of vacuum
if it completely preserves the corresponding fivebrane configuration,
but it is spontaneously broken otherwise.
 For an example of spontaneous breaking,
consider the breaking $\Z_{2n}\to \Z_2$ of the discrete chiral symmetry
of $N=1$ supersymmetric Yang-Mills theory
by a choice of vacuum configuration 
\cite{HOO,W}, or of the $N=2$ supersymmetric
$\CP^{n-1}$ sigma model in two dimensions \cite{HH}.
When a symmetry is spontaneously broken
there is a degeneracy of vacuum configurations,
as the example of $\Z_{2n}\to \Z_2$ exhibits (there are $n$
distinct vacuum configurations).

\vspace{-0.1cm}
As is well-known, the system of a flat and unrestricted fivebrane
has the six-dimensional (2,0) supersymmetry.
If the action of any of the supersymmetry generators changes the
boundary condition at infinity, the restricted system corresponding
to the four-dimensional theory no longer has that particular supersymmetry,
namely, that supersymmetry is absent from the start.
This is the case when the asymptotic boundary condition is not
holomorphic, and corresponds to an explicit supersymmetry breaking.
Examples of this kind were given in the last section of \cite{W}.
If the action of some of the supersymmetry generators preserves
the boundary condition at infinity, the restricted system
is supersymmetric under the corresponding subalgebra.
This is the case when the asymptotic boundary condition is
holomorphic, and the supersymmetric four-dimensional theory
may or may not have a supersymmetric ground state
depending on whether or not there is an everywhere holomorphic curve
obeying the boundary condition.
If the supersymmetry is spontaneously broken,
there must be massless fermions (Goldstone fermions).
The action of the supercharges
on a non-holomorphic minimal surface
generates fermionic zero modes on the brane.
When the asymptotic boundary condition is holomorphic,
these decay at infinity of the surface and therefore
correspond to fields in the four-dimensional theory. 
These can be identified with the Goldstone fermions
associated with the spontaneous supersymmetry breaking.

To summarize the main points of the above discussion: 
A supersymmetric vacuum corresponds to a fivebrane wrapping a holomorphic
complex curve.
Breaking the supersymmetry 
of the gauge theory corresponds in the brane picture to the
non-existence of
a  holomorphic curve describing the brane configuration.
A non-supersymmetric stable vacuum corresponds to a fivebrane wrapping a
nonholomorphic real two dimensional surface with minimal area,
taking into account the force between the surface and the
orientifold and the gravitational force
between the possible different components of the surface.
If supersymmetry is broken spontaneously the  
real two dimensional surface has asymptotic boundary conditions
which are holomorphic.

\subsection{Example: A Two-Dimensional Toy Model}

In the following we will present a two dimensional toy model
for supersymmetry breaking that will illustrate some of
the above discussion.
We will consider a  $U(1)$ $N=2$  supersymmetric
gauge theory in two dimensions (four supercharges) with one charged
hypermultiplet.
The theory can be obtained
by dimensional reduction of an $N=1$ supersymmetric
gauge theory in four dimensions.
The vector multiplet of the theory contains as bosonic
degrees of freedom the  gauge field $A_{\mu}$ and 
a complex scalar $\sigma$ which arises from the 
$x^2,x^3$-components of the
gauge field in the process of the dimensional reduction.
The hypermultiplet contains as bosonic degrees of freedom two
complex scalars that we will denote by $Q,\tilde{Q}$.

We can introduce a Fayet-Iliopoulos (FI) D-term, 
$-r \int d^2 x \dd^4 \theta \,V$, where
$r$ is a real FI parameter and
$V$ is the $U(1)$ vector superfield.
In addition, we introduce a complex mass term
$\int d^2 x \dd^2 \theta m Q \tilde{Q}  + {\rm c.c.}$.
The total scalar potential $U$ of the theory reads 
\be
2 e^2 U =|\sigma Q|^2 + |\sigma\tilde{Q}|^2 + |m|^2 (|Q|^2 + |\tilde{Q}|^2)
 + (|Q|^2 -|\tilde{Q}|^2 - e^2 r)^2
\ee
where $e$ is the two-dimensional gauge coupling constant. 
When $r=Q=\tilde{Q}=0$ the potential energy is zero and there is a complex
one dimensional moduli space
of supersymmetric vacua parametrized by $\sigma$.
 For $m=0,r\neq 0$ there is a complex one dimensional Higgs
branch. 
 Finally, if both $m$ and $r$ are nonzero, the potential energy
is always nonzero and supersymmetry is spontaneously broken.
If $2 e^2 r \leq -|m|^2$, the minimum of the potential is at
$(|Q|^2,|\tilde{Q}|^2)=(0,-e^2 r-|m|^2/2)$, for $-|m|^2 \leq
2 e^2 r \leq |m|^2$ it is at
$(|Q|^2,|\tilde{Q}|^2)=(0,0)$, and for 
$|m|^2 \leq 2 e^2 r $ it is at
$(|Q|^2,|\tilde{Q}|^2)=(e^2 r-|m|^2/2,0)$.
In the second case there is a family of non-supersymmetric
vacua parametrized by $\sigma$. 

Brane  configurations of $N=2$ gauge theories in two dimensions
have been studied in \cite{HH}.
In order to realize the above $U(1)$ gauge theory we consider the 
following brane
configuration in type IIA string theory.
An NS fivebrane with worldvolume coordinates
$(x^0,x^1,x^2,x^3,x^4,x^5)$, a 
NS${}^{\prime}$ fivebrane with worldvolume coordinates
$(x^0,x^1,x^2,x^3,x^8,x^9)$,
a D4 brane with worldvolume coordinates $(x^0,x^1,x^7,x^8,x^9)$ 
located between the two fivebranes, and a
 D2 brane with
worldvolume coordinates
$(x^0,x^1,x^6)$
stretched between the NS and NS' branes in the $x^6$ direction.
The theory on the worldvolume of the D2 brane in $(x^0,x^1)$
has  $N=2$ supersymmetry, gauge group $U(1)$ and one hypermultiplet
charged under the $U(1)$.
This configuration is T-dual to the $N=1$ four dimensional
configuration of  \cite{g1}.

The gauge coupling of the theory is related to the distance between
the  NS and NS${}^{\prime}$ branes in the $x^6$ direction,
$1/e^2 = (\elst/\gst){\sl \Delta}x^6$.
The FI parameter of the theory is related to the 
distance between the NS and NS${}^{\prime}$ branes in the
$x^7$ direction,
$-r= (1/\elst\gst){\sl \Delta}x^7$.
The value of the coordinates $x^2+ix^3$ of the D2 brane  
is related to $\sigma$ by  $x^2+ix^3 = \elst^2 \sigma$ and
the complex mass is related to the $x^4,x^5$ position
of the $D4$ brane via $m\elst^2 =x^4+i x^5$.

When the distance between the NS and NS${}^{\prime}$
branes in the $x^7$ direction is zero the brane configuration
preserves four supercharges and for each
value of the coordinates $x^2+ix^3$ of the D2 brane 
it corresponds to a supersymmetric vacuum. Similarly,
by breaking the D2 brane on the D4 brane we see supersymmetric
vacua for $r\neq 0,m= 0$.

%
%
%
%

Consider now the most general case $r\neq 0$ and $m\neq 0$.
There are two possible configurations for the D2 brane.
In what we will call configuration A, the D2 brane
breaks on the D4 brane, while in what we will
call configuration B it connects 
directly the 
NS and NS' branes without intersecting the D4 brane. In either
case the D2 brane no longer has its worldvolume in the
$(x^0,x^1,x^6)$ direction and therefore the brane configuration
breaks supersymmetry. The vacua correspond to the configurations
of minimal length, which is the two dimensional
analog of the minimal area
condition that we discussed in the four dimensional context.
 For configuration A, the length of the D2 brane
equals 
\be L_A=\frac{\gst}{\elst e_1^2 } + 
\sqrt{|m\elst^2|^2 + \left( \frac{\gst}{\elst e_2^2} \right)^2 }
\ee
while for configuration B it equals
\be L_B=
\sqrt{ \left( \frac{\gst}{\elst e^2} \right)^2 + 
(r \gst\elst)^2 }  .
\ee
Here, we introduced $\frac{1}{e_1^2}=\frac{\elst}{\gst}
(x^6(D4)-x^6(NS))$ and
$\frac{1}{e_2^2}=\frac{\elst}{\gst}
(x^6(NS')-x^6(D4))$.
 For large $m$, the shortest brane configuration is configuration B,
corresponding to the field theory vacua with $|Q|=|\tilde{Q}|=0$.
 For large $r$, the shortest brane configuration is configuration A,
corresponding to the field theory vacua with
$|Q|\neq 0$ or $|\tilde{Q}|\neq 0$. There is qualitative but
no quantitative agreement with field theory, 
but this is what we expect since the brane configurations
are not BPS \cite{kah}. This is already obvious by noticing that
$L_A$ depends on $e_1^2$ and $e_2^2$ which do not correspond
to parameters of field theory. Furthermore, we neglected 
the interaction between the two D2-brane components in configuration A.
In general, for non-BPS configurations, there is no reason for
this interaction to vanish. We expect that this additional
interaction will lift the degeneracy of configuration A (corresponding
to moving the component of the D2 brane between the D4 and NS' 
branes in the $8,9$-direction), which has no counterpart in field theory.

On the other hand, $L_B$ does
not depend on the parameters $e_1^2$ and
$e_2^2$ and one can indeed quantitatively
compute the energy density of the vacua with $|Q|=|\tilde{Q}|=0$
from the brane configuration. The energy density of these vacua is
given by the 
difference between $L_B$ 
and the length
$L_0=\gst/\elst e^2$
of the the supersymmetric 
brane configuration, multiplied
by the membrane tension $1/\elel^3=1/\gst\elst^3$.
This yields
$(e^2 r^2/2)\left(1+o((e^2 r \elst^2)^2)\right)$,
which converges to the field theory result
$U=e^2r^2/2$ in the
limit $\elst\to 0$ where the
the string oscillation modes decouple. 

In the non-supersymmetric brane configurations, 
the D2 brane is stretched between the NS, D4
and the NS${}^{\prime}$ branes which by themselves 
preserve supersymmetry so that the boundary conditions on the minimal
length line are supersymmetric.
This is the geometrical manifestation of the fact that 
supersymmetry is broken spontaneously and not explicitly.
  
In order to study the quantum properties of the system we have to lift
the brane configuration to M theory.
The D2 brane becomes a membrane of M theory stretched between 
pairs of fivebranes. In this framework, the theta angle $\theta$
is seen as the distance between NS and NS' fivebrane in the
$x^{10}$ direction \cite{HH} and the vacuum energy 
for configuration B becomes
$e^2|ir + \frac{\tilde\theta}{2\pi}|^2/2$ where $\tilde\theta$
is the minimum among $|{\sl\Delta}x^{10}+2\pi n|$, $n\in \Z$.
In particular, there is a discontinuity in the $\theta$-derivative
of the vacuum energy which is also what we know in field theory.

\section{$M$ Theory Description of Orientifold
via $Sp(N_c)$ Gauge Dynamics}

\newcommand{\tc}{{\it t}-configuration~}
\newcommand{\Pf}{{\rm Pf\,}}
\newcommand{\tilNc}{\widetilde{N}_c}

In this section, we study properties of the fivebrane in \MT
in a geometry which is (locally) of the type
$\R^5\times S^1\times\R^5/\Z_2$.
We construct fivebrane configurations
whose worldvolume dynamics describes some supersymmetric 
$Sp(N_c)$ gauge theories and study the properties of the fivebrane
by comparing with the structure of vacua of the corresponding
gauge theories.\footnote{
This section
is meant to be a preparation for the study of the brane realization
of a model of supersymmetry breaking with a stable non-SUSY vacuum.
Those who are mostly interested in the issue of dynamical
supersymmetry breaking can skip this section
except for the introductory part.}

In subsection 3.1,
we study the vacuum structure of
supersymmetric $Sp(N_c)$ gauge theories, including
$N=2$ SUSY QCD, $N=2$ broken to $N=1$ by a mass term for the
adjoint, and
$N=1$ SUSY QCD.
The analysis is almost the same as 
that for gauge group $SU(N_c)$
gauge group given in \cite{HOO}, but
there is an important difference: In the $N=2$ $Sp(N_c)$ theory,
the way the Higgs branch emanates from the quantum Coulomb
branch has not been determined by a field theory argument so far. 
There is a related puzzle
in the theory with finite adjoint mass, where we find an
apparent discrepancy between the dimensions of the Higgs
branches found in the $N=2$ and $N=1$ descriptions of the theory.

After reviewing some general properties of \MT on $\R^6\times
\R^5/\Z_2$ in subsection 3.2,
we construct in 3.3 fivebrane configurations 
whose worldvolume dynamics realizes the $Sp(N_c)$ gauge theories
based on the following hypothesis.

Let us denote by
``~\tc''
a configuration that looks locally like 
a single fivebrane transversely intersecting the $\Z_2$ fixed plane
at one point. More precisely, the six-dimensional worldvolume of the
fivebrane shares $\R^4$ with the
$\Z_2$ fixed plane $\R^6$ but the remaining two-dimensional part
intersects the $\Z_2$ fixed plane transversally at one point.
Then, the basic hypothesis is

{\it \tc is not supersymmetric.}

\noindent
In particular, when there are (locally) two fivebranes
intersecting the $\Z_2$ fixed plane at different points,
the two points either attract or repel each other, corresponding to
the two allowed choices of orientifold plane in the type IIA
picture.
When there is a Kaluza-Klein monopole,
we will need a modified version of the hypothesis.

Based on such a hypothesis,
we will find that the brane gives the correct
field theory results on the
structure of the moduli space of supersymmetric vacua.
Actually, this hypothesis was used in the
construction \cite{lll} of the Seiberg-Witten curve for the
$N=2$ theory with symplectic gauge group.
Here, we will see that this hypothesis is also crucial
for reproducing other aspects of $N=2$ theories
as well as properties of $N=1$ theories.
Among other things, the hypothesis plays an essential role
in reproducing the
quantum modification of the classical constraint on the meson
matrix in $N=1$ SQCD with $N_f=N_c+1$.
Moreover, based on this hypothesis, we can
determine the way the Higgs branches are emanating from
the quantum Coulomb branch in the $N=2$ theory, solving
the puzzle about the theory
with finite adjoint mass mentioned above.

\subsection{Supersymmetric $Sp(N_c)$ Gauge Theories}

\medskip
\subsection*{\sl $N=2$ SQCD}

We start with describing facts about $N=2$ supersymmetric
$Sp(N_c)$ gauge theory with $N_f$ fundamental hypermultiplets.
In $N=1$ language, it is an $Sp(N_c)$ gauge theory
with an adjoint chiral multiplet $\Phi^a_b$ and fundamental chiral
multiplets $Q^i_a$. Here $a,b,...=1,...,2N_c$ and
$i,j,...=1,...,2N_f$ are the color and flavor indices.
The matrix $J^{ab}\Phi^c_b$ is symmetric with respect to $a,c$
where $J^{ab}$ is the $Sp(N_c)$ invariant
skew-symmetric form.\footnote{In
the symplectic basis, it is represented as
the matrix
$(J^{ab})={\bf 1}_{N_c}\otimes
\left(\begin{array}{cc}
0&-1\\
1&0
\end{array}
\right)$ where ${\bf 1}_{N_c}$ is the unit matrix of size $N_c$.}
The superpotential of the theory is
\beq
W=\sqrt{2}Q^i_aJ^{ab}\Phi^c_bQ^i_c\,.
\eeq
We will consider the case $N_f<2(N_c+1)$
where the theory is asymptotically free and generates
a dynamical scale $\Lambda_{N=2}$.
The classical $U(1)$ R-symmetry group (under which
$\Phi$ carries charge $2$) is broken by an
anomaly to its discrete subgroup $\Z_{4(N_c+1)-2N_f}$
while the $SU(2)$ R-symmetry remains exact.
The flavor symmetry group is $SO(2N_f)$.

The moduli space of vacua consists of Coulomb and Higgs branches.
The Coulomb branch, where the gauge group is broken (generically)
to $U(1)^{N_c}$, is parametrized by the $N_c$ Casimirs
of $\Phi$ and is corrected by one-loop and instantons.
The quantum Coulomb branch is described by the Seiberg-Witten curve
of the form \cite{AS}
\beq
y^2=x\prod_{a=1}^{N_c}(x-\phi_a^2)^2-\Lambda_{N=2}^{4N_c+4-2N_f}
x^{N_f-1}
\label{AScurve}
\eeq
which determines the effective gauge coupling and the K\"ahler
metric. In the semi-classical region $\parallel\Phi\parallel
\gg\Lambda_{N=2}$, $\pm\phi_a$ are interpreted as the eigenvalues of
$\Phi$.

The (mixed Coulomb-)Higgs branches
of the theory were analyzed in \cite{APSh}.
They are classified by an integer $r=1,\ldots,[N_f/2]$.
The $r$-th Higgs branch has quaternionic dimension $2rN_f-(2r^2+r)$
and emanates from a $N_c-r$ dimensional complex
subspace of the Coulomb branch where there is an unbroken
gauge group $Sp(r)$.
The Higgs branches themselves are not corrected
by quantum effects, but the way they emanate from the Coulomb branch
is.
Under the na\"\i ve interpretation of $\pm\phi_a$
as the eigenvalues of $\Phi$,
the $r$-th Higgs branch emanates from the locus
where $r$ of the $\phi_a$ vanish.
This is true for the values of $r$ where the $Sp(r)$ gauge theory
with $N_f$ flavors is not asymptotic free, i.e., for $r\leq
[(N_f-2)/2]$ (all but $r=[N_f/2]$).
 For the case $r=[N_f/2]$, however, since this low energy
theory is asymptotically free and 
may be affected by strong dynamics,
it is not clear whether the Higgs branch emanates from this locus.
This problem was not solved in \cite{APSh}.
We will see that the brane gives the solution; it is true for
$r=(N_f-1)/2$ ($N_f$ odd), but is modified for
$r=N_f/2$ ($N_f$ even).

\subsection*{\sl $N=2$ Broken to $N=1$}

Let us give a bare mass $\mu$ to the adjoint chiral multiplet,
breaking $N=2$ to $N=1$ supersymmetry
\beq
W=\sqrt{2}Q^i_aJ^{ab}\Phi^c_bQ^i_c+\mu\Phi^a_b\Phi^b_a\,.
\eeq
The $U(1)$ R-symmetry modified so that $\Phi$ and $QQ$ both
carry charge $1$ is anomalously broken to $\Z_{2N_c+2-N_f}$.
 For small values of $\mu$, we can use the description in terms of the
Seiberg-Witten curve to analyze the structure of vacua.
Most of the Coulomb branch is lifted except the discrete set of points
where all the $\alpha$-cycles of the curve degenerate.
As analyzed in \cite{APSh}, the remaining vacua are in the locus where
$r=[(N_f-1)/2]$ of the parameters $\phi_a$ in (\ref{AScurve}) vanish
as well as in the locus where
$r=N_f-N_c-2$ of them vanish.
We shall call the former {\it the $\bf A$ branch root} and the latter
{\it the $\bf B$ branch root},\footnote{In
\cite{APSh}, only the $\bf B$ branch (present only for $N_f>N_c+1$)
is identified.
But this cannot be the whole thing, as is evident by considering
the case $N_c=1$ ($Sp(1)=SU(2)$), $N_f=2$ where two points in the
Coulomb branch remain.
Actually, only degenerations of the form
$y^2=x[\cdots]^2$ are considered in \cite{APSh}.
The $\bf A$ branch root corresponds to degenerations
of the form $y^2=(x-a)[\cdots]^2$, $a\ne 0$.
We thank A. Shapere for a discussion on this point.
This was also recently noticed in \cite{tera}.}
and denote the corresponding Higgs branches by the $\bf A$ and $\bf B$ 
branch respectively.
The $\bf B$ branch root is a single point
which is invariant under $\Z_{2N_c+2-N_f}$,
whereas the $\bf A$ branch root consists of $2N_c+2-N_f$ points,
namely, the $\bf A$ branch consists of $2N_c+2-N_f$ connected components;
for $N_f$ odd these are
related by $\Z_{2N_c+2-N_f}$, but for $N_f$ even they fall into two
separate orbits of
$\Z_{2N_c+2-N_f}$, each having $N_c+1-N_f/2$ components.
 Pure Yang-Mills theory 
($N_f=0$) is an exceptional case
where $N_c+1$ points related by the discrete R-symmetry
remain supersymmetric.

 For values of $\mu$ beyond $\Lambda_{N=2}$, the gauge coupling runs
below the energy scale $\mu$ as in $N=1$ supersymmetric QCD
whose dynamical scale $\Lambda_{N=1}$ is given by
\beq
\Lambda_{N=1}^{3(N_c+1)-N_f}=\mu^{N_c+1}\Lambda_{N=2}^{2(N_c+1)-N_f}
\,.
\eeq
If $\mu\gg\Lambda_{N=1}$, we can integrate out the heavy field $\Phi$.
Then, the theory at energies below $\mu$ can be considered as
$N=1$ SQCD with tree level superpotential
\beq
{\sl\Delta}W=-{1\over 2\mu}{\rm Tr}(M^2)
\,,
\eeq
where $M^{ij}=Q^iQ^j=J^{ab}Q^i_aQ^j_b$ is the meson matrix
whose components form a basis of gauge invariant chiral superfields.
The superpotential ${\sl \Delta}W$ breaks the flavor symmetry
$SU(2N_f)$ of $N=1$ SQCD to $SO(2N_f)$.
In the following, we describe the structure of vacua of this theory.

The pure $N=1$ super-Yang-Mills theory ($N_f=0$)
has $N_c+1$ massive vacua
with gaugino condensation which breaks
the discrete chiral symmetry as $\Z_{2(N_c+1)}\to\Z_2$.
These correspond to the $N_c+1$ completely degenerate curves
found in the
$N=2$ analysis which are related by the R-symmetry action.

 For $0<N_f<N_c+1$,
an Affleck-Dine-Seiberg type superpotential
is generated in the $N=1$ $Sp(N_c)$ SQCD \cite{IP}.
By a holomorphy argument as in \cite{IS,HOO},
we can show that the exact effective superpotential is obtained by
adding to ${\sl \Delta}W$ the term
\beq
W_{\it eff}=(N_c+1-N_f)\left({\Lambda_{N=1}^{3(N_c+1)-N_f}
\over \Pf M}\right)^{1/(N_c+1-N_f)}
-{1\over 2\mu}{\rm Tr}(M^2)
\,.
\eeq

 For $N_f=N_c+1$,
the moduli space of vacua of $N=1$ SQCD is modified
by quantum corrections and is given by
$\Pf M=\Lambda_{N=1}^{2(N_c+1)}$ \cite{IP}.
We therefore expect that the effective superpotential of
our model
is given by
\beq
W_{\it eff}=X(\Pf M-\Lambda_{N=1}^{2N_c+2})
-{1\over 2\mu}{\rm Tr}(M^2)\,,
\eeq
where $X$ is a chiral superfield playing the role of
a Lagrange multiplier.

 For $N_f=N_c+2$,
the exact superpotential of $N=1$ SQCD
is given by $W=-\Pf M/\Lambda^{2N_c+1}$
\cite{IP}.
The effective superpotential of our model is expected to be
\beq
W_{\it eff}=-{\Pf M\over \Lambda_{N=1}^{2N_c+1}}-{1\over 2\mu}
{\rm Tr}(M^2)\,.
\label{sconf}
\eeq

\newcommand{\mm}{\lambda}
\newcommand{\tiLambda}{\widetilde{\Lambda}}

 For $N_f>N_c+2$,
there is another
theory whose low energy physics is
equivalent to the one of the $N=1$ SQCD. 
This dual magnetic theory is an $Sp(\tilNc=N_f-N_c-2)$ gauge theory
with $2N_f$ fundamental chiral multiplets $q_i^a$
and a gauge singlet chiral multiplet $M$
(which is identified with the meson field $QQ$ of the electric theory)
which has a tree level superpotential
$W_{\it mag}={1\over \mm}M^{ij}q^a_iJ_{ab}q^b_j$.
The scale $\mm$ is needed to match the dimensions
of the fields and relates the scales $\Lambda_{N=1}$
and $\tiLambda_{N=1}$
of the electric and magnetic theories by
$\Lambda_{N=1}^{3(N_c+1)-N_f}\tiLambda_{N=1}^{3(\tilNc+1)-N_f}
=(-1)^{N_f-N_c-1}\mm^{N_f}$.
In order to study the low energy behavior of our model,
we may as well consider the magnetic theory
with the tree level superpotential
\beq
W_{\it tree}={1\over \mm}Mqq-{1\over 2\mu}{\rm Tr}(M^2)\,.
\label{magtree}
\eeq
If the rank of $M$ is maximal, ${\rm rank}\, M=2N_f$, all of
the dual quarks are massive and the low energy physics is that
of $Sp(\tilNc)$ Yang-Mills theory with the dynamical scale
$\Lambda_L$ given by $\Lambda_L^{3(\tilNc+1)}
=\Pf(M/\mm)\,\tiLambda_{N=1}^{3(\tilNc+1)-N_f}$ and an effective
superpotential given by
\beq
W_{\it eff}=(N_c+1-N_f)
\left({\Pf M\over \Lambda_{N=1}^{3(N_c+1)-N_f}}\right)^{1/(N_f-N_c-1)}
-{1\over 2\mu}{\rm Tr}(M^2)\,.
\label{maxmag}
\eeq

Proceeding as in \cite{HOO}, we find that the extremum of
$W_{\it eff}$ with ${1\over 2}{\rm rank}\,M=N_f$
is given by\\
$\bullet$ $N_f$ odd
\beq
M={\rm diag}(\overbrace{m,\ldots,m,m}^{N_f})\otimes\epsilon\,,
\label{VACo}
\eeq
$\bullet$ $N_f$ even
\beq
M=\left\{\begin{array}{l}
{\rm diag}(m,\ldots,m,m)\otimes\epsilon\,,\\
{\rm diag}(m,\ldots,m,-m)\otimes\epsilon\,,
\end{array}
\right.
\label{VACe}
\eeq
where
\beq
m=2^{-{N_c+1-N_f\over 2N_c+2-N_f}}\mu\Lambda_{N=2}\,,
\label{ei}
\eeq
or its rotation by the flavor and the R-symmetry group
$SO(2N_f)\times \Z_{2N_c+2-N_f}$.
 For $N_f$ odd, there are $2N_c+2-N_f$ distinct flavor orbits
which are related by the R-symmetry. For $N_f$ even
there are $N_c+1-N_f/2$ flavor orbits related by
the R-symmetry from the first type
of solutions of (\ref{VACe}) and the same number of orbits
related by $\Z_{2N_c+2-N_f}$
from the second type, and in total there are $2N_f+2-N_f$
distinct flavor orbits.
Each of the flavor orbits is the homogeneous space
$SO(2N_f,\C)/H_{\C}$ where $H_{\C}$ is the complexification of
the unbroken subgroup of $SO(2N_f)$
at the diagonal solution (\ref{VACo}),
(\ref{VACe}). For (\ref{VACo}) and the first type of (\ref{VACe}),
the unbroken subgroup is $H=U(N_f)$, whereas for the second type of
(\ref{VACe}) it is some other subgroup of the same dimension as $U(N_f)$.
In any case, the homogeneous space has complex dimension
$N_f(N_f-1)$.

 For small values of $\mu$,
this result must be consistent with the analysis based on the
$N=2$ description.
 From the $\Z_{2N_c+2-N_f}$ symmetry breaking pattern,
these are to be identified with the
$\bf A$ branch in the $N=2$ analysis.
Recall that the $\bf A$ branch emanates from the locus where
$r=[(N_f-1)/2]$ of $\phi_a$'s are vanishing. At first sight this
suggests that
the $\bf A$ branch is the $r=[(N_f-1)/2]$ Higgs branch.
 For odd $N_f$ it is consistent because
the dimension of the $r=(N_f-1)/2$ Higgs branch has complex dimension
$4rN_f-2(2r^2+r)=N_f(N_f-1)$.
However, there is a puzzle for even $N_f$:
The complex dimension of the $r=[(N_f-1)/2]=(N_f-2)/2$ Higgs branch
is $4rN_f-2(2r^2+r)=N_f(N_f-1)-2$ which is smaller by $2$ than the
above result. We will
solve this puzzle in the brane picture.

These are the only supersymmetric vacua for $N_f\leq N_c+1$.
 For $N_f\geq N_c+2$, there is another
branch with ${1\over 2}{\rm rank}M<N_f$.

 For $N_f=N_c+2$, we can consider (\ref{sconf}) to be
the exact superpotential which is valid for all 
values of the
ranks of $M$.
There is a unique solution of lowest rank of the extremum equation.
It is
\beq
M=0\,.
\eeq
This new solution corresponds to the $\bf B$ branch of
the $N=2$ analysis since the $\bf B$ branch is expected to be the
$r=N_f-N_c-2=0$ Higgs branch which has dimension zero.

 For $N_f>N_c+2$, we cannot consider (\ref{maxmag}) 
as the exact superpotential for non-maximal rank$M$.
The classical flat direction, or the extremum locus of
(\ref{magtree}), is described as
\beq
{\rm rank}M\leq 2\tilNc\,,\quad M^2=0\,,
\label{Spco}
\eeq
and $q\cdot q=(\mm/\mu)M$.
In this direction,
at least $2N_f^{\prime}=2N_f-2\tilNc$
of the dual quarks are massless.
Thus, the low energy theory is the $Sp(\tilNc)$ gauge theory
with massless quarks $q_{\it low}^i$,
$i=1,\ldots,2N_f^{\prime}$
(the flavor is larger for rank$M<2\tilNc$).
Note that $N_f^{\prime}-(\tilNc+2)=2(N_c+1)-N_f$ is positive
because our starting point was an asymptotically free $N=2$ theory.
Thus, the origin $q_{\it low}^i=0$ of the moduli space of the
low energy theory is not lifted.
This shows that the quantum moduli space contains the
classical flat direction (\ref{Spco}) as one of its
branches.\footnote{
This is in contrast with the case of $SU(N_c)$ gauge theory
with $N_f>N_c+1$ flavors and a heavy adjoint
in which the analogous branch is described at low energy
by the $SU(\tilNc)$ gauge theory with $\tilNc$ massless quarks.
The quantum modification of the moduli space
of this low energy theory \cite{Seiberg94}
induces a modification of this branch
\cite{H,HOO}.}
One can show as in \cite{H} that there are no other branches.
We note that the constraint (\ref{Spco}) is the same as the one
defining the Higgs branch of $N=2$ $Sp(\tilNc)$ gauge theory
with $N_f$ flavors \cite{APSh}.
Thus, this branch is identified as the $\bf B$
branch of the $N=2$ analysis which is expected
to be the $r=\tilNc$ Higgs branch because
$r=N_f-N_c-2=\tilNc$ of the $\phi_a$ are vanishing at the root.

\subsection*{\sl The $\mu\to\infty$ Limit}

As $\mu\to\infty$ keeping $\Lambda_{N=1}$ finite,
because
\beq
(\mu\Lambda_{N=2})^{2N_c+2-N_f}
=\mu^{N_c+1-N_f}\Lambda_{N=1}^{3N_c+3-N_f},
\eeq
the branch with maximal rank ${1\over 2}{\rm rank}M=N_f$
runs away to infinity for $N_f<N_c+1$,
whereas for $N_f\geq N_c+1$
it remains finite and constitutes a submanifold
of dimension $N_f(N_f-1)$ of the moduli space of SQCD.

The lower rank branch which is present for $N_f\geq N_c+2$
remains finite in the $\mu\to\infty$ limit
and constitutes a submanifold $M^2=0,
{1\over 2}{\rm rank}M\leq \tilNc$ of dimension
$2(2\tilNc N_f-(2\tilNc^2+\tilNc))$
of the moduli space
${1\over 2}{\rm rank}M\leq N_c$ of SQCD.

\subsection{$M$ Theory Realization of Orientifold Four-Plane}

In the next subsection, we will construct a
\MT fivebrane configuration whose worldvolume dynamics
describes at long distances the $Sp(N_c)$ gauge theory
with $N=2$ or $N=1$ supersymmetry.
In the weakly coupled Type IIA regime, the configuration
involves an orientifold four-plane (O4-plane).
Here, we briefly explain the \MT realization of an O4-plane
in Type IIA string theory.

In Type IIA string theory, there are two types of O4-plane
which are classified by the RR charges \cite{Polchinski}.
One type carries D4-brane charge $-1$ if it is counted before taking the
$\Z_2$ quotient (and therefore, $-1/2$ after the quotient),
while
the other carries D4-brane charge $+1$.
When there are $2n$ D4-branes close to the $\Z_2$ fixed plane
of the former type
there is a $SO(2n)$ gauge symmetry in the common directions
of the branes (which is generically broken to $U(1)^n$
spontaneously), 
while the gauge symmetry is $Sp(n)$ for the latter type.
 For this reason,
we shall call the former {\it SO-type}
and the latter {\it Sp-type}.
In other words, an open string ending on D4-branes at an $SO$-type
O4-plane carries $SO(2n)$ Chan-Paton factors
while an open string ending on D4-branes at an $Sp$-type
O4-plane carries $Sp(n)$ Chan-Paton factors.

An orientifold of Type IIA on $\R^5\times\R^5/\Z_2$
is obtained from \MT on the orbifold $\R^5\times\R^5/\Z_2\times S^1$
by taking the limit
where the radius of $S^1$ is small.
\MT on $\R^6\times T^5/\Z_2$ was studied in \cite{Worb,DM} 
and it was concluded from the local cancellation of gravitational anomaly
that each of the $\Z_2$ fixed plane must carry fivebrane charge
$-1$ (in the unit before quotient).
Since a fivebrane wrapped on $S^1$ becomes the D4 brane in the
Type IIA limit, this shows that the O4-plane which is identified as the
$\Z_2$ fixed plane compactified on $S^1$ carries D4-brane charge $-1$.
Namely, the O4-plane obtained in this way is of $SO$-type.

Then, what is the \MT interpretation of the $Sp$-type O4-plane?
Here we propose that it can be obtained by
putting two fivebranes at the $\Z_2$ fixed point set
$\R^5\times S^1$, and then
taking the $\Z_2$ quotient
such that the degrees of freedom corresponding to
the motion of the fivebranes away from the $\Z_2$ fixed points
are frozen.
The D4-brane charge of the $\Z_2$ fixed plane
in the Type IIA limit is now $-1+2=+1$,
which agrees with the $Sp$-type O4-plane.
 For further test of the idea, let us
put $2n$ additional D4-branes parallel and close to this
$\Z_2$ fixed plane.
Then, the light degrees of freedom on the worldvolume
in the common 4+1 dimensions
are created by open strings ending on the $2n+2$ D4-branes
where $+2$ are the two D4-branes corresponding
to the two fivebranes
stuck at the $\Z_2$ fixed point.
Since the motion of the two D4-branes are frozen,
degrees of freedom coming from the open strings with both ends
at these two D4-branes are killed.
Then, it is easy to see that the remaining light degrees of freedom
have the same mass spectrum as those associated with
the $N=2$ $Sp(n)$ gauge theory in 4+1 dimensions.
Note however that we do not see directly
 how the above proposed definition of 
 $Sp$-type O4-plane indeed yields $Sp$ gauge group.

In general, we must assume the existence of
such a $\Z_2$ quotient that freezes
the motion of the two fivebranes.
However, in what follows we consider a configuration with
NS 5-branes transversal to the O4-plane
in the Type IIA regime. The presence of such NS 5-branes
naturally freezes the two D4 branes within the standard
$\R^5/\Z_2$ orbifold of \MT (which, in the absence
of the fivebranes, would yield the $SO$-type
O4-plane in the Type IIA limit).
This follows from the basic hypothesis
that a \tc is not supersymmetric, as we will
now see.

\subsection{The Fivebrane Configuration}

\subsection*{\sl Type IIA Configuration}

\begin{figure}[htb]
\begin{center}
\epsfxsize=4in\leavevmode\epsfbox{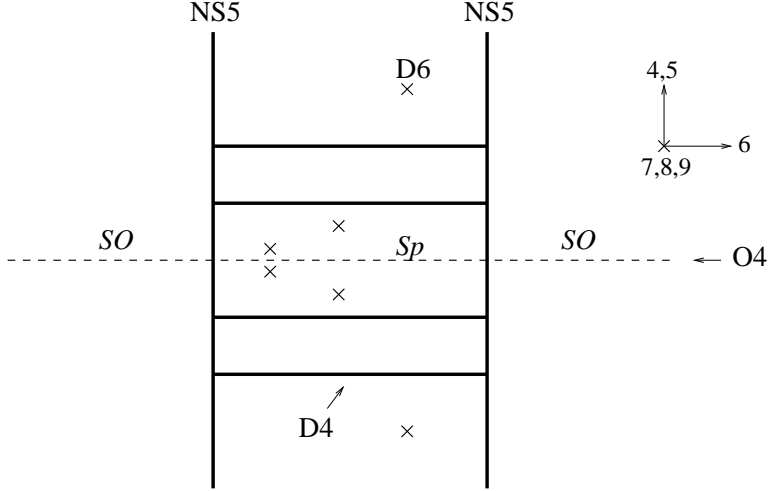}
\end{center}
\caption{Type IIA Configuration}
\label{IIA}
\end{figure}

We can realize
the $Sp(N_c)$ gauge theories as considered in the first subsection
by configurations in a Type IIA orientifold on $\R^5\times \R^5/\Z_2$
which involve (before the $\Z_2$ quotient)
two NS 5-branes, $2N_c$ D4-branes parallel to
the O4-plane, and $2N_f$ D6-branes.
We denote the time and space coordinates
by $x^0$ and $x^1,\ldots,x^9$,
where the $\Z_2$ acts as the sign change of $x^{4,5,7,8,9}$
so that the O4-plane spans the 01236 directions.
The NS 5-branes are separated in the $x^6$ direction, the
D4-branes are stretched between them, and the D6-branes,
spanning the 0123789 directions, are located between them.
In the configuration corresponding to the $N=2$ theory,
the two NS 5-branes are parallel and span the 012345 directions.
Giving a mass to the adjoint corresponds to rotating one of them
in the 45-89 direction \cite{Barbon},
and sending the mass to infinity corresponds
to taking the right angle limit, in which the rotated NS 5-brane
spans the 012389 directions. See figure \ref{IIA}.

In any case, the O4-plane is separated into three pieces by the two
NS 5-branes. In order to obtain $Sp(N_c)$ gauge theory
on the worldvolume of the D4 branes,
we want the O4-plane to be of $Sp$-type (D4-brane charge $+1$)
in the part between the two NS
5-branes. 
Since the flavor group of the $N=2$ theory is $SO(2N_f)$,
the other two parts of the O4-plane should be of $SO$-type
(D4-brane charge $-1$) in the $N=2$ configuration \cite{ejs}.
This has been proven using a world-volume computation in
\cite{ggkk}. The other two parts of the O4-plane must then also
be of $SO$-type for all other configurations
that can be obtained by a rotation of the $N=2$ configuration
(and possibly by other continuous deformations).

\subsection*{\sl $M$ Theory Realization of O4-NS5 System}

We first give an \MT interpretation of the simplest configuration
of two NS 5-branes intersecting with the O4-plane of the type
described right above. We consider the case of parallel NS5-branes
spanning the 012345 directions.
It is described by a configuration of a single fivebrane
in \MT on the orbifold $\R^5\times \R^5/\Z_2\times S^1$
and it is located at the origin in the 789 directions.
We denote by $x^{10}$ the coordinate of the circle $S^1$,
$x^{10}\equiv x^{10}+2\pi$.
Let us introduce the complex coordinates
\beqa
t&=&\exp\left(-{x^6\over R}-ix^{10}\right)\,,\\
v&=&(x^4+ix^5)\times {\rm const}\,,
\eeqa
where $R$ is the radius of $S^1$.
The $\Z_2$ acts on these coordinates as $t\to t, v\to -v$.

There are two regions of the fivebrane
worldvolume
with large values of $v$
corresponding to the two NS 5-branes.
Since the D4-brane charge jumps by $+2$ when crossing the left NS 5-brane
from left to right,
the corresponding region of the fivebrane 
worldvolume
behaves as $t\sim v^2$ at large $v$.
Similarly, the charge jumps by $-2$ for the right NS5-brane and therefore
the corresponding region behaves as $t\sim v^{-2}$ at large $v$.
A general fivebrane invariant under the $\Z_2$ action $v\to -v$
which satisfies these boundary conditions is given by
\beq
t^2-(v^2+c)t+\zeta=0\,,
\label{O4NS}
\eeq
where $c$ and $\zeta$ are parameters with $\zeta\ne 0$.
The coefficient of the $v^2$ term can be taken to be $1$
by a rescaling
of $t$ which is a shift of the origin of the $x^{6,10}$ directions.
At first sight, there are two light degrees of freedom
corresponding to the two moduli. Obviously, $\zeta$ is related to
the distance between the two NS 5-branes and therefore can be considered
as frozen (the kinetic energy for the variation of $\zeta$ diverges
because of the infinite volume of the fivebrane).
However, we still have one parameter $c$ which does not appear in the
Type IIA configuration we want.
Without the $\Z_2$ quotient this modulus would have
a finite kinetic energy, as it is equivalent with the
modulus of the Coulomb branch of $N=2$ super-Yang-Mills theory
with gauge group $SU(2)$.

Now, this is the time where we can use the basic hypothesis
to eliminate this degree of freedom.
 For generic values of $c$, the fivebrane (\ref{O4NS}) and the
O4-plane $v=0$ intersect transversely at two points, each of which is a
{\it t}-configuration.
In the case
\beq
c=2\sqrt{\zeta}\,,
\eeq
(\ref{O4NS}) looks $(t-\sqrt{\zeta})^2\sim v^2$
and we see that the two intersection points collide. 
If we accept the hypothesis
that a \tc is not supersymmetric
and two such intersection points attract each other,
then the configuration is stable only in the case $c=2\sqrt{\zeta}$.
In this way, we can eliminate the degrees of freedom corresponding to
varying $c$.\renewcommand{\thefootnote}{\fnsymbol{footnote}}
\footnote[3]{There are actually two choices of the sign of $\sqrt{\zeta}$.
They are related by a coordinate change
and therefore are equivalent.
We will see some related phenomena in the following
as we shall note in footnotes
with the same symbol $\ddag$.}
\renewcommand{\thefootnote}{\arabic{footnote}}
 From now on we will always assume that the intersection points
attract each other for an $SO-SP-SO$-type O4-plane.

\subsubsection{$N=2$ Configurations}

\subsection*{\sl Coulomb Branch}

We next review the \MT fivebrane configuration
describing the Coulomb branch of $N=2$
$Sp(N_c)$ SQCD with $N_f$ massless flavors following \cite{lll}.
In the Type IIA picture where all the $2N_f$ D6-branes are sent to
$x^6=+\infty$, there are $2N_f$ semi-infinite D4-branes
ending on the right NS5-brane from the right.
The $\Z_2$ invariant curve satisfying a suitable boundary condition
is of the form $t^2-(v^{2N_c+2}+u_1 v^{2N_c}+\cdots)t+
\zeta v^{2N_f}=0$.
The requirement of absence of a \tc fixes one of the coefficients $u_i$ of
the polynomial $v^{2N_c+2}+u_1 v^{2N_c}+\cdots$, and it is easy to see
that the curve for $N_f\ne 0$ is of the form
\beq
t^2-v^2B_{N_c}(v^2)t+\Lambda^{4N_c+4-2N_f}_{N=2}v^{2N_f}=0\,,
\label{N=2curve}
\eeq
where $B_{N_c}=v^{2N_c}+\sum_{i=1}^{N_c}u_i v^{2N_c-2i}$.
 For $N_f=0$, in order to avoid the {\it t}-configuration,
$v^2B_{N_c}(v^2)$ must be replaced by
$v^2B_{N_c}(v^2)+
2\Lambda_{N=2}^{2N_c+2}$.\renewcommand{\thefootnote}{\fnsymbol{footnote}}
\footnote[3]{Note that $2\Lambda^{2N_c+2}_{N=2}$ here can be replaced
by $-2\Lambda^{2N_c+2}_{N=2}$.
This means that the moduli space of holomorphic curves
obeying the required condition splits into two disconnected
components.
However, the change of the sign of $\Lambda_{N=2}^{2N_c+2}$ can be
absorbed by a coordinate change, and the two connected components
are actually equivalent. Thus, we can choose only one component
as far as we consider small excitations around the supersymmetric
vacua.}
\renewcommand{\thefootnote}{\arabic{footnote}}
If we identify $\Lambda_{N=2}$ and the $u_i$'s with
the dynamical scale and Casimirs of $\Phi$,
this is the same as the Seiberg-Witten curve (\ref{AScurve})
describing the Coulomb branch of the $N=2$ gauge theory.

\subsection*{\sl Higgs Branch}

We next present a description of the $N=2$ Higgs branch in the brane
picture.
Because of our insufficient understanding of the strong coupling dynamics
of Type IIA string theory in the presence of an orientifold four-plane
and D6-branes, the description will be far from complete.\footnote{
This is in contrast with the case of an O6-plane and D6-branes where
there is a complete geometric description in $M$ theory \cite{Sei}.
This leads to a nice description of the $N=2$ Coulomb and
Higgs branches in terms of the fivebrane \cite{HOV}
(see also \cite{LL}). The results obtained for $N=2$ theories can
actually be more straightforwardly obtained using an O6-plane.}
However, by imposing some rule which generalizes our basic hypothesis,
we can give a fairly reasonable description in terms of the fivebranes
in $M$ theory on a certain orbifold.
In particular, we can find a location of the $r=[N_f/2]$ Higgs branch
on the quantum Coulomb branch.

\begin{figure}[htb]
\begin{center}
\epsfxsize=4in\leavevmode\epsfbox{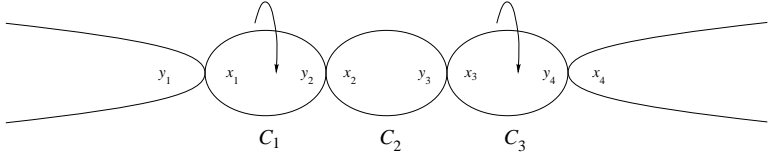}
\end{center}
\caption{The $\Z_2$ action for $N_f=2$ case}
\label{Z2}
\end{figure}

We consider taking the $\Z_2$ orbifold projection of
\MT on a Taub-NUT geometry which represents the $2N_f$ D6-branes.
In the case when all the D6-branes are at $v=0$,
with a particular choice of complex structure the Taub-NUT space
is described by
\beq
xy=\Lambda_{N=2}^{4N_c+4-2N_f}v^{2N_f}
\eeq
where $x$ and $y$ are complex coordinates.
We can resolve the $A_{2N_f-1}$ singularity at $x=y=v=0$
by a complex surface covered by $2N_f$ patches
with coordinates $(y_i,x_i)$ ($i=1,\ldots,2N_f$) that are
glued by $(y_{i-1},x_{i-1})=(y_i^2x_i,y_i^{-1})$ and are mapped to
$(y,x,v)$ as
$y=\Lambda^{2N_c+2-N_f} y_i^ix_i^{i-1}$,
$x=\Lambda^{2N_c+2-N_f}y_i^{2N_f-i} x_i^{2N_f+1-i}$,
and $v=y_ix_i$.
The singular point $x=y=v=0$ has been blown up to
$2N_f-1$ $\CP^1$ cycles $C_i$
defined by $y_i=x_{i+1}=0$ for $i=1,\ldots,2N_f-1$.
The $\Z_2$ group acts on $(x,y,v)$ as $\to (x,y,-v)$ or on the
coordinates of the $i$-th patch as
$y_i\to (-1)^{i-1}y_i$, $x_i\to (-1)^i x_i$
(see figure \ref{Z2}).
We see that the $\Z_2$ fixed plane has split into two infinite and
$N_f-1$ $\CP^1$ cycles, namely, the infinite cigar at $x_1=0$,
the cycles $C_{i}$ for even $i$, and another infinite cigar at
$y_{2N_f}=0$.

The configuration of the fivebrane is given by
\beq
x+y=v^2B_{N_c}(v^2)=v^2\prod_{a=1}^{N_c}(v^2-\phi_a^2)\,.
\label{N=2curve2}
\eeq
This is obviously the same as (\ref{N=2curve}) under the
identification $y=t$.

\begin{figure}[htb]
\begin{center}
\epsfxsize=3.5in\leavevmode\epsfbox{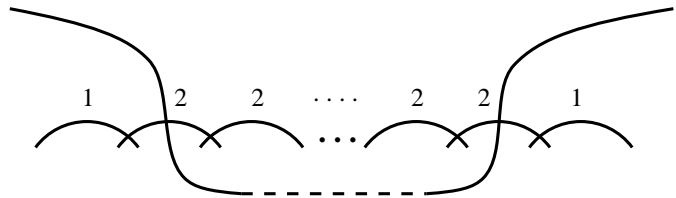}
\end{center}
\caption{A Generic Point on the Coulomb Branch; The number above the
$\CP^1$ components stands for the multiplicity.}
\label{Coulomb}
\end{figure}

Let us first look at a generic point of the Coulomb branch where
$B_{N_c}(v^2)$ is not zero at $v=0$. As in \cite{HOO},
we see that the fivebrane wraps the $\CP^1$ cycles
$C_1,C_2,\ldots,C_{2N_f-2},C_{2N_f-1}$ with multiplicity
$1,2,\ldots,2,1$. Also, an infinite component intersects
$C_2$ at one point and $C_{2N_f-2}$ at one point
(see figure \ref{Coulomb}).
Recall that the $\Z_2$ acts on $\vec{w}\simeq (x^7,x^8,x^9)$
as $\vec{w}\to-\vec{w}$ at the same time.
Only from the requirement of the $\Z_2$ invariance of the configuration,
there is nothing to prevent the two components wrapped on
$C_i$, $2\leq i\leq 2N_f-2$, to be separated in the opposite direction
of $\vec{w}$,
which would correspond
to some modulus of the worldvolume theory; from the transformation
property under the $SU(2)$ R-symmetry
it would be a hypermodulus. But this does not agree with the
field theory knowledge; In $N=2$ $Sp(N_c)$ SQCD, no Higgs branch
emanates from a generic point of the Coulomb branch.
Therefore, these degrees of freedom must be eliminated by some
mechanism if the worldvolume theory is close to the $Sp(N_c)$
gauge theory at all.
We remark that the infinite component
intersects $C_2$ and $C_{2N_f-2}$ at one point each.
An obvious way to eliminate these extra moduli is to generalize
our basic hypothesis as follows.

{\it Suppose that the fivebrane intersects one or two different $\CP^1$ cycles
of the $Z_2$ fixed plane at two distinct points. 
Then, the configuration is
supersymmetric only if the two distinct intersection points are connected by
one or more $\CP^1$ components  of the fivebrane.}

\noindent
Indeed, under the constraint that the intersection points at $C_2$ and
$C_{2N_f-2}$ be connected by a series of $\CP^1$ components,
there is no $\Z_2$ invariant way to move the components wrapped on
$C_i$ away from $\vec{w}=0$. 

Let us next consider the case where $B_{N_c}(v^2)$ has a zero at $v=0$ of
order $2r$,
\beq
B_{N_c}(v^2)\sim  c\, v^{2r}+{\rm higher},
\eeq
where $c$ is a non-zero constant.
We first consider the range $r<[N_f/2]$.
Then, there are $\CP^1$ components wrapped on $C_1,\ldots,C_{2N_f-1}$
with multiplicity
$1,2,3,\ldots,2r+1,2r+2,\ldots,2r+2,2r+1,\ldots,3,2,1$
(there are $2N_f-4r-3$ $\CP^1$ cycles wrapped with multiplicity  
$2r+2$), and also there is an infinite curve intersecting 
the first and last $\CP^1$ of multiplicity $2r+2$.
If we impose the generalized basic hypothesis,
from each of the $\CP^1$'s of multiplicity $2r+2$
only $r$ pairs of components can be separated in the $\vec{w}$
direction.
Thus, the number of separable pairs of $\CP^1$ components are
\beqa
&&1+1+2+2+\cdots+r+r+(2N_f-4r-3)\times r
+r+r+\cdots+1+1\nonumber\\
&&=4{r(r+1)\over 2}+(2N_f-4r-3)r
=2rN_f-(2r^2+r),
\eeqa
which agrees with the dimension of the $r$-th Higgs branch.

 For the case $r=(N_f-1)/2$ ($N_f$ odd),
the multiplicities of $\CP^1$ cycles are
$1,2,\ldots,2r,2r+1,2r,\ldots,2,1$
(there is no cycle with multiplicity $2r+2$)
and the infinite curve intersects the $\CP^1$
of multiplicity $2r+1$. The number of separable
pairs of $\CP^1$ components are
$4r(r-1)/2+3r=N_f(N_f-1)/2$ which is the dimension of
the $r=(N_f-1)/2$ Higgs branch.

 For higher $r$, the pattern of degeneration of the curve is the same
as in the case of $r=(N_f-1)/2$ (for $N_f$ odd) or $r=(N_f-2)/2$
(for $N_f$ even).

So far, we have identified the $r=1,2,\ldots,[(N_f-1)/2]$
Higgs branches.
Then where is the $r=N_f/2$ Higgs branch (in the case $N_f$ even)?
Does it disappear in the quantum theory?

There is actually a subtlety in the case $r=(N_f-2)/2$ ($N_f$ even).
In this case the number of $\CP^1$'s with multiplicity $2r+2$
is $2N_f-4r-3=1$. Thus, there is a possibility that the two
intersection points of the infinite curve with the $\CP^1$'s coincide.  
In that would happen, there
would be no restriction from the generalized basic hypothesis
concerning the separation of pairs of the $\CP^1$ components
of the fivebrane.
In order to see whether this is indeed what happens, let us look at the $N_f$-th
patch with coordinate $(y_{N_f},x_{N_f})$.
Equation (\ref{N=2curve2}) looks near $x=y=v=0$ as
\beq
y_{N_f}^{N_f}x_{N_f}^{N_f-1}\left(x_{N_f}^2
-{c\over\Lambda_{N=2}^{2N_c+2-N_f}}x_{N_f}+1\right)=0.
\eeq
This shows in particular that the $\CP^1$
of multiplicity $2r+2=N_f$ is the locus $y_{N_f}=0$.
 For generic values of $c$ the infinite component
(described by $x_{N_f}^2
-{c\over\Lambda_{N=2}^{2N_c+2-N_f}}x_{N_f}+1=0$) intersects the
$\CP^1$ of multiplicity $2r+2$ at two different points,
and therefore at least one $\CP^1$ component of the fivebrane must be
at $\vec{w}=0$ by the basic hypothesis.
However, for
\beq
c=\pm 2\Lambda_{N=2}^{2N_c+2-N_f},
\eeq
the two intersection points collide. Then
the basic hypothesis imposes no restrictions,
and we have one additional pair of $\CP^1$
components that can be separated, and
the total number of movable pairs becomes
\beq
2rN_f-(2r^2+r)+1={N_f(N_f-1)\over 2}\,.
\eeq
which is the dimension of the $r^{\prime}=N_f/2$ Higgs branch.
We conclude that we have indeed found the
$r=N_f/2$ Higgs branch.
Namely, {\it the $r=N_f/2$ Higgs branch, which emanates classically
from the locus where $N_f/2$ of the $\phi_a$'s vanish,
emanates in the quantum theory from the locus where
only $(N_f-2)/2$ of them vanish
 and the product of
non-zero $\phi_a^2$'s is $\pm 2\Lambda^{2N_c+2-N_f}$.}
This generalizes the quantum
splitting of the Higgs branch root $u=0\to \pm 2\Lambda^2$
in the $SU(2)$ gauge theory with $N_f=2$ flavors.

There remains a question of understanding the theory at the root
of the $r=N_f/2$ Higgs branch. As noted before, an $Sp(N_f/2)$ gauge
theory with $N_f$ flavors is asymptotically free and becomes strongly
coupled at the scale $\Lambda_{N=2}$. Is the theory at the root
a new non-trivial fixed point?
At least for the $N_c=1$ case, we know another description of the
theory (as an $SU(2)$ gauge theory) and we know that the theory at the
root is just an $N=2$ free QED with two electrons.
This may suggest that in general the gauge group splits
into $U(1)$ and $Sp(N_f/2-1)$ and the theory flows to a direct product
of a free QED and the
conformal field theory of $Sp(N_f/2-1)$ with $N_f$ flavors.

\subsubsection{$N=1$ Configurations}

\subsection*{\sl Rotation by Finite Angles}

Next, we construct the fivebrane configuration for the theory
with an adjoint mass $\mu$.
This corresponds to rotating, say, the left NS 5-brane
in the 45-89 plane while keeping fixed the right
NS 5-brane \cite{Barbon}.
The left (right) NS 5-brane corresponds to the asymptotic region
with large $v$ where the curve behaves as $t\sim v^{2N_c+2}$
($t\sim v^{2N_f-2N_c-2}$).
In terms of the complex coordinate
\beq
w=(x^8+ix^9)\times{\rm const}\,,
\eeq
the boundary condition at the left infinity $t\sim v^{2N_c+2}$
reads
\beq
w\,\sim\,\mu v\,, 
\label{bc}
\eeq
while it is $w\sim 0$ at the right infinity $t\sim v^{2N_f-2N_c-2}$.
We require that the configuration is invariant
under the rotations $U(1)_{45}$ and $U(1)_{89}$
in the 45 and 89 planes which are identified with
$U(1)$ R-symmetries under which $\Phi$ carries charge
$2$ and $0$ respectively.
Under the group $U(1)_{45}\times U(1)_{89}$,
the coordinates and parameters
are charged as $v:(2,0)$, $w:(0,2)$, $\mu:(-2,2)$,
$t:(4N_c+4,0)$, $x:(4N_c+4,0)$, and
$\Lambda_{N=2}^{2N_c+2-N_f}:(4N_c+4-2N_f,0)$.
The combination ${1\over 2}U(1)_{45}+{1\over 2}U(1)_{89}$
makes $\mu$ invariant, and its $\Z_{2N_c+2-N_f}$ subgroup
makes $\Lambda_{N=2}^{2N_c+2-N_f}$ invariant.
The latter is interpreted as the
non-anomalous $\Z_{2N_c+2-N_f}$ R-symmetry
of the field theory. 
As in \cite{HOO},
from the invariance under the $U(1)$ symmetries,
we can conclude that the projection of the rotated curve is the same
as the curve before rotation.
Then, the $w$ values of the rotated curve
can be considered as a function on a original, fixed curve.

Let us first consider the case where the curve has a single
infinite component.
When the curve is compactified at the two points with $v=\infty$,
$w$ has a simple pole at one infinity by (\ref{bc})
and hence the compactified curve is
necessarily (birational to) $\CP^1$, and the rotated curve
is a cylinder which is globally parametrized by $w$.
Proceeding as in \cite{HOO}, we can determine the allowed form of
$t$ and $v$ as functions of $w$.
The only difference from \cite{HOO} 
is that here we require the curve
to be invariant under the $\Z_2$: $(t,v,w)\to (t,-v,-w)$,
and we find
\beqa
v&=&\mu^{-1}w^{-1}(w^2-M^2)\,,
\label{finvw}\\
t&=&\mu^{-2N_c-2}w^{2N_c+2-2N_f}(w^2-M^2)^{N_f}\,.
\label{fintw}
\eeqa
By requiring that this projects to a curve in the $t$-$v$ plane
of the form (\ref{N=2curve}), we find that for $N_f>0$
$M$ must satisfy the equation
\beq
M^{4N_c+4-2N_f}=(-1)^{N_f}(\mu^2\Lambda_{N=2}^2)^{2N_c+2-N_f}\,.
\eeq
There are $2N_c+2-N_f$ solutions for $M^2$.
The $\Z_{2N_c+2-N_f}$ rotational symmetry is completely
broken by any of such
$M^2$'s for $N_f$ odd while it is broken to
$\Z_2$ for $N_f$ even, since its generator acts on $M^2$ as
\beq
M^2\to \e^{4\pi i\over 2N_c+2-N_f} M^2\,.
\eeq
 For $N_f=0$, we find $M^{2N_c+2}=
(\mu^2\Lambda_{N=2}^2)^{N_c+1}$ and there are $N_c+1$ solutions
for $M^2$ related by the $\Z_{2N_c+2}$ R-symmetry.

These properties are the same as
the structure of vacua with maximal rank solutions for the
meson vev in the $N=1$ superpotential analysis,
which are identified as the $\bf A$ branch of the $N=2$ analysis.
Indeed, one can show that the function $B_{N_c}(v^2)$ defining the
projected curve in the $t$-$v$ plane is of the form
$B_{N_c}(v^2)=c\,v^{2r}+\cdots$(higher order terms), $c\ne 0$,
where $r\geq (N_f-1)/2$ for $N_f$ odd and $r=(N_f-2)/2$
for $N_f$ even which is the property of the $\bf A$ branch.
This can be seen as follows:
Note that as $w\to\pm M$, both $v\to 0$ and $t\to 0$, because
$t\sim v^{N_f}$ as can be seen from
(\ref{finvw})-(\ref{fintw}).
On the other hand, equation (\ref{N=2curve}) implies
\beq
t\simeq {c\over 2}v^{2r+2}\pm\sqrt{\left({c\over 2}v^{2r+2}\right)^2-
\Lambda_{N=2}^{4N_c+4-2N_f}v^{2N_f}}\,.
\label{temp1}
\eeq
The two statements are consistent only when $r\geq (N_f-2)/2$.
If $N_f$ is even, equations (\ref{finvw})-(\ref{fintw})
show that $t$ is a single valued function of $v$ near $v\sim 0$
(for odd $N_f$, it is two-valued because the choice
of $w=+M$ or $-M$ affects the sign of $t$).
This is possible only if the two terms in the square root of
(\ref{temp1}) cancel, namely, when $r=(N_f-2)/2$
and $c=\pm2\Lambda_{N=2}^{2N_c+2-N_f}$.

Recall now that there was a puzzle
in the field theory analysis for even $N_f$
concerning the dimension of the moduli space:
The $\bf A$ branch emanates from the locus of the
Coulomb branch where $r=(N_f-2)/2$ of the $\phi_a$
vanish, and from this we expected that this branch is
the $r=(N_f-2)/2$ Higgs branch which has complex
dimension $N_f(N_f-1)-2$.
However,
$N=1$ analysis shows that the dimension is $N_f(N_f-1)$.
The solution of the puzzle is that the $\bf A$ branch
is actually the $r=N_f/2$ Higgs branch which, in the $N=2$ theory,
emanates from
the locus where $r=(N_f-2)/2$ of the $\phi_a$ are vanishing
and the product of the non-zero $\phi_a^2$ is $\pm
2\Lambda^{2N_c+2-N_f}$,
as we have seen.

To summarize, the branch of vacua that remain after turning on a 
mass for the adjoint that we considered so far
is identified with the branch in the
$N=1$ superpotential analysis where the meson has maximal rank, and with the
$\bf A$ branch of the $N=2$ analysis.
{\it In the $N=2$ limit, 
the $\bf A$ branch is the $r=[N_f/2]$ Higgs branch}.

Next, we consider the case where the curve is factorized
so that the two infinities are separated.
When such a factorization occurs, we just have
to rotate the component including the left infinity by
$w=\mu v$.
It is easy to see that the factorization of (\ref{N=2curve})
is unique and is given by
\beq
(t-v^{2N_c+2})(t-\Lambda_{N=2}^{4N_c+4-2N_f}v^{2N_f-2N_c-2})=0\,.
\eeq
By expanding this in $t$, we see that
$v^2B_{N_c}(v^2)=v^{2N_c+2}+
\Lambda_{N=2}^{4N_c+4-2N_f}v^{2N_f-2N_c-2}$.
In particular,
this factorization is possible only for $N_f\geq N_c+2$.
Also, this shows that the curve belongs to the $r=N_c-N_f-2$
Higgs branch root.
Therefore, we can identify this as the $\bf B$ branch of
the $N=2$ analysis which is identified with the branch
in the $N=1$ analysis with non-maximal rank meson vev's.

\subsection*{\sl The Limit $\mu\to \infty$}

\newcommand{\tily}{\widetilde{y}}
\newcommand{\tilt}{\widetilde{t}}

We consider taking the $\mu\to\infty$ limit.
As in the case of
$SU(N_c)$ gauge group \cite{HOO}, we need to 
rescale the coordinate as $\tily(=\tilt)=\mu^{2N_c+2}t$
in order to not let the configuration run away. 
The complex structure of the space-time (before the
$\Z_2$ quotient)
is described by $\tily x=\Lambda_{N=1}^{6N_c+6-2N_f}v^{2N_f}$,
where
$\Lambda_{N=1}^{3N_c+3-N_f}=\mu^{N_c+1}\Lambda_{N=2}^{2N_c+2-N_f}$.
$\Lambda_{N=1}^{3N_c+3-N_f}$ is invariant under the
$\Z_{2N_c+2-2N_f}$
($\Z_{2N_c+2}$) subgroup of $U(1)_{45}$ ($U(1)_{89}$).

 For $N_f=0$, the limit of one of the $N_c+1$ curves
is given by
\beqa
&&\tilt=w^{2N_c+2}\,,\\
&&vw=-\Lambda_{N=1}^3\,,
\eeqa
and the limits of the other $N_c$ curves
are given by the action of the $\Z_{2N_c+2}$ symmetry.
This in particular exhibits the chiral symmetry breaking
$\Z_{2N_c+2}\to\Z_2$.

 For $0<N_f<N_c+1$, the curve becomes infinitely elongated
because $M^2\to\infty$ in the limit. This is consistent with
the absence of supersymmetric vacua for this number of flavors
in the field theory.

 For $N_f\geq N_c+1$, the limit exists. It contains two infinite
components and several
$\CP^1$ components. The infinite components --- $C_L$ and $C_R$ ---
are given by
\beq
C_L\left\{
\begin{array}{l}
\tily=(w^2-M^2)^{N_c+1},\\
v=0,
\end{array}
\right.
\qquad
C_R\left\{
\begin{array}{l}
x=v^{2N_c+2},\\
w=0,
\end{array}
\right.
\eeq
where $(M^2)^{N_c+1}=(-\Lambda_{N=1}^4)^{N_c+1}$ for $N_f=N_c+1$
while $M^2=0$ for $N_f>N_c+1$.
The $\CP^1$ components are of multiplicity
$2N_c+2,\ldots, 2N_c+2,2N_c+1,2N_c,\ldots,3,2,1$ from left to right
(the number of $2N_c+2$'s is $2N_f-2N_c-2$).
 For $N_f\geq N_c+2$,
the component $C_L$ intersects the left-most $\CP^1$
at $w=0$ and
the component $C_R$ intersects
the last $\CP^1$ transversely with multiplicity $2N_c+2$.
 For $N_f=N_c+1$, both do not intersect any of the $\CP^1$'s
at $w=0$.

\subsection*{\sl Configuration for $N=1$ SQCD}

After the limit $\mu\to\infty$, the curve acquires new deformation directions
for $N_f\geq N_c+1$. The component $C_R$ is still rigid but $C_L$ can be deformed
as $\tily=(w^2-M_1^2)\cdots (w^2-M_{N_c+1}^2)$, $v=0$. 
However, there is a constraint for the allowed values
of $M_i^2$ which is required by the basic hypothesis.

Let us first consider the case of $N_f=N_c+1$.
In this case, the component $C_R$
can be described as $\tily=\Lambda_{N=1}^{4N_c+4}$,
$w=0$, and hence it intersects the $\Z_2$ fixed point set
at one point $(x,\tily,v,w)=(0,\Lambda_{N=1}^{4N_c+4},0,0)$.
This is a \tc and it is not supersymmetric by itself
according to the basic hypothesis.
The intersection point attracts the other infinite component $C_L$
so that $C_L$ intersects the $\Z_2$ fixed point set at the
same point. This is equivalent to requiring
\beq
M_1^2\cdots M_{N_f}^2=(-1)^{N_c+1}\Lambda_{N=1}^{4N_c+4}\,.
\eeq
This corresponds to the quantum modified constraint
$\Pf M=\Lambda_{N=1}^{2N_c+2}$
on the meson matrix.

In the case $N_f\geq N_c+2$,
the component $C_R$ intersects transversely with the $2N_c+2$nd
$\CP^1$ from the right.
 From the generalized basic hypothesis,
there must be a chain of $\CP^1$ components that connects
$C_R$ with $C_L$. Since $C_R$ is at $w=0$, all such
$\CP^1$ components should also be at $w=0$ and in particular
$C_L$ must intersect the left-most $\CP^1$ at $w=0$.
This means that some of the $M_i^2$'s must be zero.
This corresponds to the classical
constraint ${\rm rank} M\leq 2N_c$ which is also the full constraint
in the quantum theory.

One can also see the agreement in the dimension of the moduli space.
 For $N_f\geq N_c+2$
there is a constraint for the motion of $\CP^1$'s
due to the generalized basic hypothesis, while
there is no constraint on the $\CP^1$ motion for $N_f=N_c+1$.
In either case the dimension is the sum of $N_c$ from the variation of
the $M_i$ and $2(1+1+\cdots +N_c+N_c+(2N_f-2N_c-2)\times N_c)$
from the $\CP^1$ motion, and is in total
\beq
N_c+4{N_c(N_c+1)\over 2}+2(2N_f-2N_c-2)\times N_c=
4N_fN_c-(2N_c^2+N_c),
\eeq
which agrees with what we know from field theory.

\section{Izawa-Yanagida-Intriligator-Thomas Model}

\newcommand{\MUV}{M_{\rm UV}}

\subsection{Field Theory Analysis}

\subsection*{\sl The Model}

In \cite{IY,IT}, a  non-chiral model of dynamical
supersymmetry breaking with stable non-supersymmetric vacua
was given. Classically it is defined as an $N=1$ supersymmetric
$Sp(N_c)$ gauge theory with $2N_f=2(N_c+1)$ fundamental
chiral multiplets
$Q^i_a$ ($a=1,\ldots,2N_c$, $i=1,\ldots,2N_f$)
and a gauge singlet chiral multiplet $S_{ij}$ which is
an anti-symmetric tensor with respect to the flavor
indices $i,j$.
The tree level superpotential is given by
\beq
W_{\it tree}=\lambda S_{ij}Q^iQ^j\,,
\label{IYIT}
\eeq
where $Q^iQ^j=J^{ab}Q^i_aQ^j_b$
form the basis of gauge invariant chiral superfields,
\footnote{$J^{ab}$ is the skew-symmetric matrix of size
$2N_c\times 2N_c$ which is preserved by the group $Sp(N_c)$.
Likewise, we will use $J^{ij}$ to denote the $Sp(N_f)$
invariant matrix of size $2N_f\times 2N_f$.
In a symplectic basis,
the matrix $J=(J^{ij})$ is given by $J={\bf 1}_{N_f}\otimes\epsilon$
where ${\bf 1}_{N_f}$ is the identity matrix of size $N_f$ and
$\epsilon=\left(\begin{array}{cc}
0&-1\\
1&0
\end{array}
\right).
$
}
and $\lambda$ is the Yukawa coupling constant.
The coupling (\ref{IYIT})
is chosen so that the $SU(2N_f)$ flavor symmetry
is preserved.
In this paper, we will consider a deformation of this
model by a linear term in $S$ in the superpotential
\beq
W_{\it tree}=\lambda S_{ij}Q^iQ^j-m J^{ij}S_{ij}\,,
\label{dIYIT}
\eeq
which breaks the flavor symmetry to $Sp(N_f)$.
We shall denote this theory also by the IYIT model
\cite{IY,IT}, although the deformation by the linear term
was not considered there.

The exact effective superpotential of the theory is given by
\beq
W_{\it eff}=X(\Pf M-\Lambda^{2N_c+2})+\lambda S_{ij}M^{ij}
- m J^{ij}S_{ij}\,,
\eeq
where 
$M^{ij}=-M^{ji}$
is the meson matrix corresponding to $Q^iQ^j$
and $\Lambda$ is the dynamical scale of the $Sp(N_c)$ gauge interaction.
$X$ in the first term is a chiral superfield which plays the role
of a Lagrange multiplier.
The variation of $W_{\it eff}$ with respect to $X$, $S$ and $M$
yields
\beqa
\Pf M&=&\Lambda^{2N_c+2}\,,
\label{QMC}\\
M&=&{m\over \lambda}J\,,
\label{MC}\\
S&\propto& M^{-1}\,.
\label{SM}
\eeqa
The first equation is the quantum modified constraint
which represents the $Sp(N_c)$ gauge dynamics
in the case of $N_f=N_c+1$ \cite{IP}.
The first two conditions (\ref{QMC}) and (\ref{MC})
are consistently satisfied only if
$\Pf(mJ/\lambda)=\Lambda^{2N_c+2}$, namely
\beq
\left({m\over\lambda}\right)^{N_c+1}=\,\,\Lambda^{2N_c+2}\,.
\label{Pfm}
\eeq
Therefore, if the condition (\ref{Pfm}) is not satisfied,
the supersymmetry is spontaneously broken \cite{IY,IT}.

If (\ref{Pfm}) is satisfied,
the supersymmetry is not broken. Indeed,
the meson matrix has a fixed vev $M=mJ/\lambda$
satisfying the quantum modified
constraint (\ref{QMC}), and there is a complex one-dimensional
flat direction for the values of $S$ because the condition
(\ref{SM}) implies
$S_{ij}\propto J_{ij}$.

\medskip
\subsection*{\sl Existence Of Non-supersymmetric Stable Vacuum}

Let us analyze the model with
$(m/\lambda)^{N_c+1}\ne\Lambda^{2N_c+2}$
where the supersymmetry is dynamically broken. For simplicity,
we consider the original model of \cite{IY,IT} where $m=0$,
but the same analysis applies also to the case $m\ne 0$.

 For large values of $S$, the fundamental chiral multiplets are heavy
and the theory at energies below $\lambda S$ is the pure $N=1$
super-Yang-Mills theory whose dynamical scale $\Lambda_L$
is given by $\Lambda_L^{3N_c+3}=\Pf(2\lambda S) \Lambda^{2N_c+2}$. 
The superpotential of the low energy theory
is given by
\beq
W_L=(N_c+1)\Lambda_L^3=(N_c+1)\,\Pf(2\lambda S)^{1/(N_c+1)}
\Lambda^2.
\eeq
Around the locus $S_{ij}=\sigma J_{ij}$,
all the components of $S_{ij}$ other than
$\delta\sigma J_{ij}$ are massive,
and the effective superpotential with respect to $\sigma$
is given by
\beq
W_L=2(N_c+1)\lambda \sigma\Lambda^2\,.
\eeq
We see that the supersymmetry is indeed broken in this direction
$\partial W_L/\partial\sigma\sim\lambda\Lambda^2\ne 0$.

The scalar potential is thus given by
\beq
U=g^{\sigma\bar \sigma}
\left|{\partial W_L\over\partial\sigma}\right|^2
=g^{\sigma\bar \sigma}|2(N_c+1)\lambda\Lambda^2|^2\,,
\label{POT}
\eeq
where $g^{\sigma\bar\sigma}$ is the inverse of the K\"ahler
metric $g_{\sigma\bar \sigma}$ with respect to $\sigma$.
At large values of $\sigma$, we can evaluate the metric
$g_{\sigma\bar\sigma}$ by perturbation theory.
The dominant correction is due to the one loop of the $Sp(N_c)$
fundamental chiral multiplets, and the corrected metric
is given by
\beq
g^{1-{\rm loop}}_{\,\sigma\bar\sigma}
=1-{N_c\over 4\pi^2}|\lambda|^2
\log\left|{\lambda\sigma\over \MUV}\right|\,,
\label{oneloop}
\eeq
where $\MUV$ is the ultra-violet cut off.
As we increase $\sigma$, $g^{1-{\rm loop}}_{\,\sigma\bar\sigma}$
decreases, and hence the scalar potential (\ref{POT})
grows.
\footnote{Strictly speaking, the one-loop correction
(\ref{oneloop}) is reliable for large $\sigma$ such that
$|\lambda\sigma|$ is almost close to the cut off $\MUV$.
However, a simple renormalization group argument \cite{AM}
shows that this
potential rise persists for much smaller values of $\sigma$.}
Since the scalar potential grows at large values of
$|\sigma|$, there must be a minimum at some smaller values of
$\sigma$. This ensures the existence of a stable vacuum.
\footnote{More rigorously, we must show that the potential grows in
{\it all} possible directions in the $S_{ij}$ space (at large values).
This seems also to be true. Consider for example the case where
$(S_{ij})={\rm diag}(\sigma_1,\ldots,\sigma_{N_f})
\otimes \epsilon$ in which first $r$ $\sigma_i$'s are much smaller
than the last $N_f-r$: $\sigma_1\sim\sigma_2\sim\cdots\sim\sigma_r\sim
\sigma_{\it small}\ll\sigma_{\it large}\sim\sigma_{r+1}\sim\cdots
\sim\sigma_{N_f}$. Then, the superpotential
of the effective
theory at energies below $\lambda\sigma_{\it large}$ is given by
$$
W_L\sim (\sigma_{\it small})^{r/(N_c+1)}\,.
$$
It gives the run-away potential with respect to
$\sigma_{\it small}$ since
$\partial W_L/\partial \sigma_{\it small}\sim
(\sigma_{\it small})^{r/(N_c+1)-1}$
is a negative power of $\sigma_{\it small}$
and the potential
cannot be made to grow by a perturbative correction.
Thus, $\sigma_{\it small}$ runs away to larger values until
it becomes comparable to $\sigma_{\it large}$.
We thank H. Murayama for explaining us this argument.}
If there is no singularity in the K\"ahler metric,
the vacuum has finite energy and breaks supersymmetry.

The sign $-$ (minus) of the one-loop correction (\ref{oneloop})
is essential for the potential growth at large $\sigma$.
This would not have been the case if, say,
we had gauged the $SO(2N_f)$ subgroup of the flavor group
as we will do in the next section.

\medskip
\subsection*{\sl Perturbation By Quadratic Term}

\newcommand{\hM}{\widehat{M}}
\newcommand{\lsim}{\stackrel{\displaystyle <}{\sim}}

 For later use, we also consider a deformation of the superpotential 
(\ref{dIYIT}) by a quadratic term in $S$:
\beq
\Delta W={1\over 2\mu}S_{ij}S^{ij}
\label{quad}
\eeq
where the flavor indices are raised and lowered by the
Kronecker delta $\delta^{ij}$ which is an $SO(2N_f)$ invariant.
This breaks the flavor symmetry further down to
$Sp(N_f)\cap SO(2N_f)=U(N_f)$.
The effective superpotential of the deformed theory is given by
\beq
W_{\it eff}=X(\Pf M-\Lambda^{2N_c+2})
+\lambda S_{ij}M^{ij}- m J^{ij}S_{ij}
+{1\over 2\mu}S_{ij}S^{ij}
\,,
\eeq
The extremum condition reads as
\beqa
&&\Pf M\,=\,\Lambda^{2N_c+2}
\label{PfC2}\\
&&M^{ij}-{m\over\lambda}J^{ij}\,=\,{1\over\lambda\mu} S^{ij}
\label{MJS}\\
&&\lambda S_{ij}+{1\over 2} X\,\Pf M\, (M^{-1})_{ij}\,=\,0\,.
\eeqa
The last two equations yield
\footnote{Here and in what follows,
the obvious multiplication by a Kronecker delta is abbreviated.
 For example, the matrix symbol $M$ means
$M^{ij}$ as well as $M^i_{\,\,j}$ (or $M_{ij}$) and what it means
should be clear from the context.}
\beq
M-{m\over \lambda} J\,\propto\,M^{-1}\,.
\label{MJM}
\eeq
This in particular implies that the anti-symmetric matrix $M$
commutes with $J$, which shows that it can be expressed as
\beq
M=A\otimes\epsilon+B\otimes {\bf 1}_2
\eeq
where ${\bf 1}_2$ is the $2\times 2$ identity matrix and
$A$ and $B$ are $N_f\times N_f$ symmetric and antisymmetric
matrices respectively, ${}^tA=A$, ${}^tB=-B$.
In other words, $M$ can be mapped to an $N_f\times N_f$
matrix $\hM=A-iB$ which transforms in the (complexified)
adjoint representation
of the flavor group $U(N_f)$.
The equations (\ref{PfC2}) and (\ref{MJM}) then yield
\beqa
&&\det\hM\,=\,\Lambda^{2N_c+2}\\
&&\hM^2-{m\over\lambda}\hM\,\propto\,{\bf 1}_{N_f}\,.
\eeqa
As in \cite{HOO}, we can show that $\hM$ is diagonalizable
for generic values of $m$,\footnote{
Non-diagonalizable solutions are possible only for
$(m/2\lambda)^{N_c+1}=\Lambda^{2N_c+2}$ where
there are solutions with Jordan blocks of size two.}
and there are at most two kinds of eigenvalues.
Proceeding as in \cite{HOO}, we see that there are solutions
parametrized by
$r=0,1,\ldots,[N_f/2]$, where the solution of the type $r$
is such that $r$ of the eigenvalues of $\hM$ are $M_+$
and $N_f-r$ of them are $M_-$ where
\beq
(M_+)^r(M_-)^{N_f-r}=\Lambda^{2N_c+2}\,,\qquad
M_++M_-={m\over\lambda}\,.
\eeq
A general solution can be obtained from the complexified flavor
rotation of a diagonal solution, and hence the
moduli space of type $r$ solutions is the homogeneous space
$GL(N_f,\C)/(GL(r,\C)\times GL(N_f-r,\C))$
which is of complex dimension $2r(N_f-r)$.

The vev of $S_{ij}$ is determined by $M$ through (\ref{MJS}),
$S=\lambda\mu(M-mJ/\lambda)$:
\beq
S={\rm diag}\Bigl(\overbrace{\sigma_+,\ldots,\sigma_+}^{r},
\overbrace{\sigma_-,\ldots,\sigma_-}^{N_f-r}\Bigr)\otimes \epsilon\,
\label{Seq}
\eeq
where $\sigma_{\pm}=\lambda\mu(M_{\pm}-m/\lambda)=-\lambda\mu M_{\mp}$
and thus
\beq
(\sigma_+)^{N_f-r}(\sigma_-)^{r}=(-\lambda\mu\Lambda^2)^{N_c+1}\,,
\qquad
\sigma_++\sigma_-=-\mu m\,.
\label{sigmapm}
\eeq
As we take the limit $\mu\to\infty$,
either $\sigma_+$ or $\sigma_-$ diverges, and hence
$S$ runs away to infinity unless $r=0$ and $\sigma_-=$ finite.
This is possible if and only if the parameter $m$ satisfies the
constraint $(m/\lambda)^{N_c+1}=\Lambda^{2N_c+2}$.
This is consistent with
the constraint (\ref{Pfm}) for the existence of supersymmetric
vacuum for the theory without the quadratic term (\ref{quad}).
The vacuum obtained in this way has the vevs
$M=mJ/\lambda$, $S=0$.
Other vacua with general $S$
can be obtained by tuning $m$ to approach $\Lambda^2$
as $m=\lambda\Lambda^2-\sigma/\mu$. Then
the $r=0$ solution
converges in the limit $\mu\to\infty$  to
$M=mJ/\lambda$, $S=\sigma J$.

Let us consider the case
$(m/\lambda)^{N_c+1}\ne \Lambda^{2(N_c+1)}$
where the supersymmetric vacua run away to infinity $S\to\infty$
as $\mu\to\infty$.
We have seen in the $\mu=\infty$ case that there is
a (presumably non-supersymmetric) vacuum at a small value of $S$.
Is there a local minimum for finite $\mu$
which approaches this vacuum in the $\mu\to\infty$ limit?
Let us again look at the region of large $S$ with
$S_{ij}=\sigma J_{ij}$.
The effective superpotential is given by
\beq
W_{\it eff}=2(N_c+1)\left(\sigma(\lambda\Lambda^2-m)
-{\sigma^2\over 2\mu}\right)\,.
\eeq
By taking into account the one-loop correction to the
K\"ahler metric of $S_{ij}$
for large values of $\sigma$ so that $|\lambda\sigma|$
is almost close to the ultra-violet cut off $\MUV$,
the effective scalar potential is given by
\beq
U= 4(N_c+1)^2
\left|\,(\lambda\Lambda^2-m)-{\sigma\over\mu}\,\,\right|^2
\left(1-{N_c\over4\pi^2}|\lambda|^2\log
\left|{\lambda\sigma\over\MUV}\right|\right)^{-1}
\label{scalarpot}
\eeq
In the direction of the $\sigma$-plane where $\sigma$ has the
same phase as $\mu(\lambda\Lambda^2-m)$,
the first factor has a negative slope
in the region $\sigma<\mu(\lambda\Lambda^2-m)$ while
the second factor has a positive slope around $|\lambda\sigma|\lsim
\MUV$. Therefore, whether $U$ is rising or not is a consequence of
the conflict of the two factors. It turns out that it is rising
near $|\lambda\sigma|\lsim \MUV$
(and hence there is a local minimum at smaller values)
when $\mu$ is large enough so that
\beq
\lambda\mu(\lambda\Lambda^2-m)>
\left(\,{8\pi^2\over N_c|\lambda|^2}+1
\,\right)\MUV\,.
\eeq

\begin{figure}[htb]
\begin{center}
\epsfxsize=2.5in\leavevmode\epsfbox{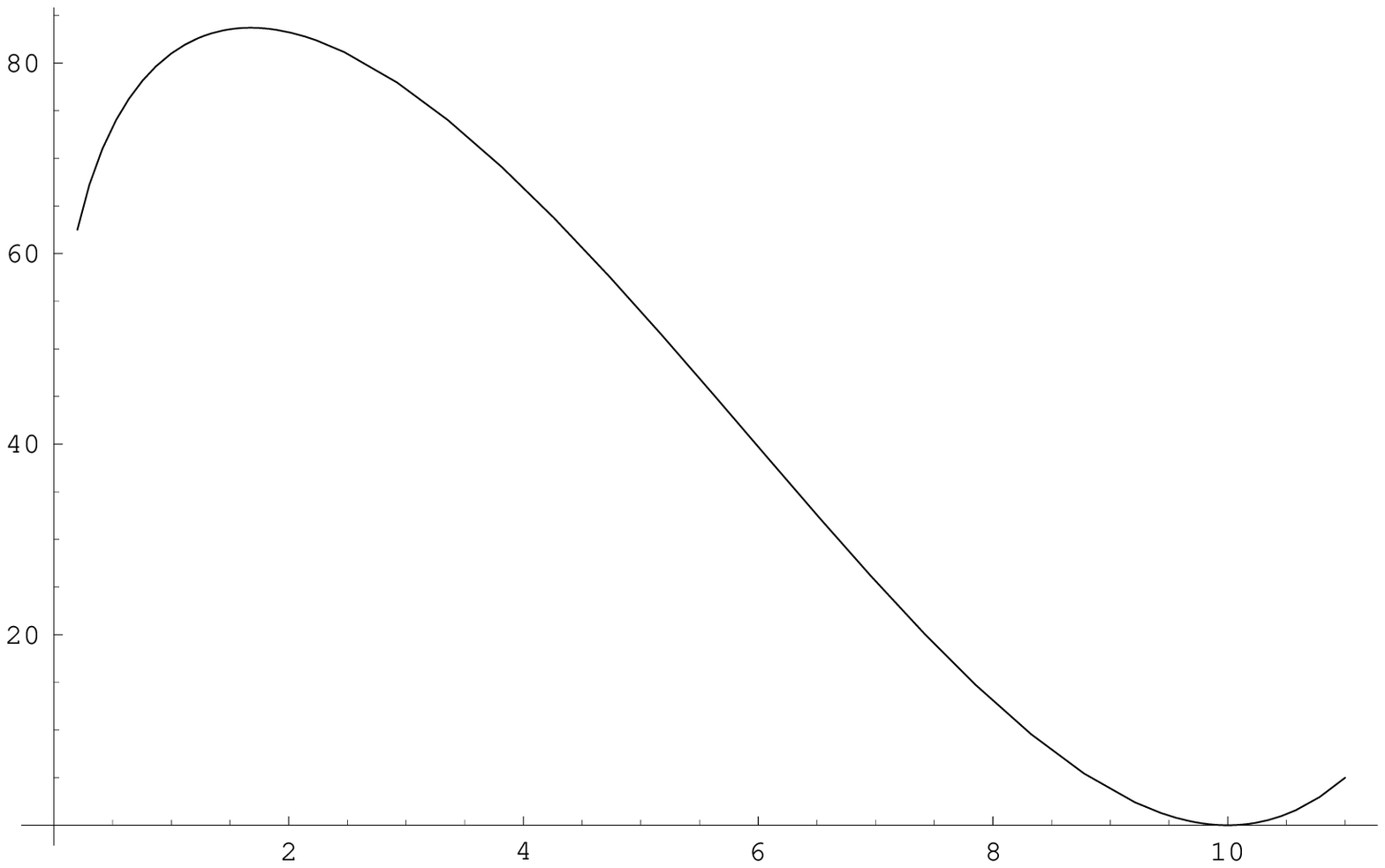}
\end{center}
\caption{Graph of $(10-s)^2/(1-{1\over 3}\log s)$}
\label{graph1}
\end{figure}

 For illustration, we present in figure \ref{graph1}
the graph of the scalar potential
$U$ (\ref{scalarpot}),
up to an overall normalization,
for the case
$\MUV=1$, $\lambda\mu(\lambda\Lambda^2-m)=10$,
$N_c|\lambda|^2/4\pi^2=1/3$, where the horizontal axis parametrizes
$s=\lambda\sigma$. Note that the function (\ref{scalarpot})
makes sense as scalar potential
only in the region below the cut-off $\lambda\sigma\leq \MUV$,
and hence only the region $s\leq 1$ is important 
although we continued it to larger values.

\subsection{Brane Realization of the Model}

We construct a brane configuration corresponding to the 
IYIT model.
We first construct the configuration for the model
perturbed by the quadratic term ${\sl \Delta}W=S^2/\mu$
which contains supersymmetric vacua,
and then consider the limit $\mu\to\infty$.

\subsection*{\sl Configuration for Finite $\mu$}

\newcommand{\tilL}{\widetilde{\Lambda}}

We first note that the tree-level superpotential
of the theory
\beq
W_{\it tree}=\lambda S_{ij}Q^iQ^j- m J^{ij}S_{ij}
+{1\over 2\mu}S_{ij}S^{ij}
\,,
\eeq
is the same as the tree-level superpotential
of the magnetic dual of the $Sp(\tilNc)$
gauge theory (which we shall call the electric dual)
with $2N_f$ quarks $q_i$ of bare mass $m$
with a quartic superpotential $(qq)^2/2\mu$.
Although $\tilNc=N_f-N_c-2=-1$,
we can still make use of this fact as a guide to
construct the brane configuration corresponding to the
perturbed IYIT model\footnote{Recently, this procedure was justified 
in \cite{H} by showing that the configuration constructed in this way
correctly reproduces the IYIT model in the type IIA limit.}.
Namely, we first consider
$N_f$, $N_c$ to be 
in the region where the electric-magnetic duality holds
and then take the limit $\tilNc\to -1$
after construction of the configuration.
We then read off the boundary condition satisfied by the curves
obtained in this way, and identify all the curves
with the same boundary condition. As evidence we will show how
to reproduce some of the results of section~4.1 from these
brane configurations.

The meson field $q_iq_j$ of the electric dual
is identified with $S_{ij}$. The tree-level superpotential
$-m qq+(qq)^2/2\mu$ shows that this theory
is obtained from the $N=2$ $Sp(\tilNc)$ SQCD with $N_f$ flavors
by giving a bare mass $\mu$ to the adjoint chiral multiplet.
Note that the dynamical scale $\tilL$ of the electric dual is
given by
$\Lambda^{3(N_c+1)-N_f}\tilL^{3(\tilNc+1)-N_f}=(-1)^{\tilNc+1}
\lambda^{-N_f}$ and it is related to
the dynamical scale $\tilL_{N=2}$
of the high energy $N=2$ theory by
$\tilL^{3(\tilNc+1)-N_f}
=\mu^{\tilNc+1}\tilL_{N=2}^{2(\tilNc+1)-N_f}$.

The configuration for $\mu=0$ is given by
\beq
t^2-\left( v^2B_{\tilNc}(v^2)+2\tilL_{N=2}^{2\tilNc+2-N_f}m^{N_f}
\right)t+\tilL_{N=2}^{4\tilNc+4-2N_f}(v^2+m^2)^{N_f}=0\,.
\label{n2c}
\eeq
In the standard way, we can find the configuration
for finite $\mu$. The result is
\beqa
vw&=&\mu^{-1}(w^2+w_+w_-),\\
t&=&\mu^{-2\tilNc-2}w^{2\tilNc+2-2N_f}(w^2-w_+^2)^r
(w^2-w_-^2)^{N_f-r},
\eeqa
where $r=0,1,\ldots,[N_f/2]$ and $w_+, w_-$ are solutions of
\beqa
&&w_++w_-\,=\,i\mu m,\\
&&w_+^{N_f-r-(\tilNc+1)}w_-^{r-(\tilNc+1)}
\,=\,
(-1)^{\tilNc+1}i^{N_f}(\mu\tilL_{N=2})^{N_f-2(\tilNc+1)}.
\eeqa

At this stage, we put $\tilNc=-1$. Then the relation among the
dynamical scales is
$\tilL_{N=2}=\tilL=\lambda\Lambda^2$, and the configuration is
given by
\beqa
vw&=&\mu^{-1}(w^2+w_+w_-),
\label{vweq}\\
t&=&w^{-2N_f}(w^2-w_+^2)^r
(w^2-w_-^2)^{N_f-r},
\label{tweq}
\eeqa
where $r=0,1,\ldots,[N_f/2]$ and
\beq
w_++w_-=i\mu m,\qquad
w_+^{N_f-r}w_-^{r}=
(i\lambda\mu\Lambda^2)^{N_f}.
\label{wpm}
\eeq
Note that
\beq
v^2+m^2=\mu^{-2}w^{-2}(w^2-w_+^2)(w^2-w_-^2)\,.
\eeq
 From this and the equation (\ref{tweq}),
we see that
the curve passes through the $A_{N_f-1}$ singularity
at $v=\pm im$, $t=x=0$ in such a way that
there are $\CP^1$ components with multiplicity
$1,2,\ldots,r-1,r,\ldots,r,r-1,\ldots,2,1$ (the number of $r$'s
is $N_f-2r+1$)
in the resolved surface.

As far as $w_{\pm}\ne 0$, these curves satisfy the following
boundary conditions.
The curve extends to infinity in two directions:
\beqa
&&v\sim \mu^{-1} w,\,\,\, t\sim 1,\,\,\,w\sim\infty,
\label{bc1}\\
\mbox{and}&&w\sim 0,\,\,\,
t\sim (\lambda\Lambda^2)^{-2N_F}v^{2N_f},\,\,\,v\sim\infty.
\label{bc2}
\eeqa
Also
\beq
t=0\,\,{\rm implies}\,\, v=\pm im.
\label{bc3}
\eeq
It is easy to identify all
holomorphic curves satisfying this boundary condition.
Those with a single infinite component
are the curves given above and similar ones
but with the second equation of (\ref{wpm}) 
replaced by $w_+^{N_f-r}w_-^{r}=-
(i\lambda\mu\Lambda^2)^{N_f}$.
\renewcommand{\thefootnote}{\fnsymbol{footnote}}
\footnote[3]{The two types of solutions are related
by a change of the sign of $\Lambda^{2N_f}$. Thus, the presence of
such copies is related to
the doubling phenomena which we encountered before. In the present case,
there is no obvious way to relate these curves by a coordinate
change and it is not clear whether they can be considered as
equivalent.
However, as the presence of
such copies does not make any change in the discussion
of dynamical supersymmetry breaking,
we simply ignore this doubling, by assuming that there is a way to
show the equivalence to the curves (\ref{vweq})-(\ref{wpm}).}
\renewcommand{\thefootnote}{\arabic{footnote}}

\begin{figure}[htb]
\begin{center}
\epsfxsize=4in\leavevmode\epsfbox{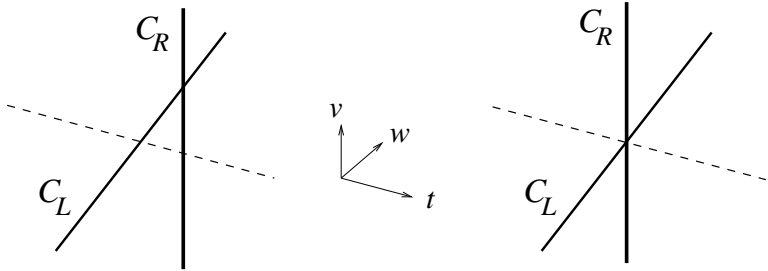}
\end{center}
\caption{The configuration (\ref{facto}) near the $\Z_2$ fixed plane
for $m^{N_f}\ne (\lambda\Lambda^2)^{N_f}$ (left) and $m^{N_f}=
(\lambda\Lambda^2)^{N_f}$ (right). Note that the left one is a
{\it t}-configuration.}
\label{figIYIT}
\end{figure}

In addition to these, there is a solution consisting of two infinite
components. It is given by
\beq
C_L\left\{
\begin{array}{l}
t=1,\\
v={\mu}^{-1} w,
\end{array}
\right.
\qquad
C_R\left\{
\begin{array}{l}
t=(\lambda\Lambda^2)^{-2N_f}(v^2+m^2)^{N_f},\\
w=0.
\end{array}
\right.
\label{facto}
\eeq
It is easy to see that this curve
does satisfy the boundary condition given
above.
However, for generic values of $m$,
the two components intersect
the $\Z_2$ fixed plane $v=w=0$ at different points,
$C_L$ at $t=1$ while $C_R$ at $t=(m/\lambda\Lambda^2)^{2N_f}$
(see figure \ref{figIYIT}),
and therefore the configuration is
not supersymmetric by the basic hypothesis.
The two intersection points coincide only when
$m^{N_f}=(\lambda\Lambda^2)^{N_f}$, 
and only then can the configuration be considered as defining
a supersymmetric vacuum.
Actually, in the case
$m^{N_f}=(\lambda\Lambda^2)^{N_f}$, one of the $r=0$ solutions
of (\ref{vweq})-(\ref{tweq})
has $w_-=0$ and factorizes into two components,
and is identical to the above solution.

As in \cite{HOO}, we can
interpret $w_{\pm}$ as the eigenvalues of the matrix $S_{ij}$
by the R-symmetry and the flavor symmetry breaking pattern.
Indeed, there is a precise correspondence of
the values for each $r$ (compare (\ref{tweq}) with (\ref{Seq})
and (\ref{wpm}) with (\ref{sigmapm})):
\beq
\sigma_{\pm}=iw_{\pm}\,.
\eeq
We also note that there are $r(N_f-r)$ $\CP^1$ components at $v=im$
together with their mirror images at $v=-im$
and this is consistent with the fact that the complex
dimension of the $r$-th branch is equal to $2r(N_f-r)$.
The extra solution present
in the case $m^{N_f}=(\lambda\Lambda^2)^{N_f}$
can be identified with the supersymmetric vacuum with
$S=0$.

\subsection*{\sl The Supersymmetric Configurations for $\mu=\infty$}

Let us take the $\mu\to\infty$ limit by keeping fixed $\lambda$,
$\Lambda$ and $m$.
The configuration with a single infinite component
goes away from the region with finite $vw$
and becomes infinitely elongated in the $x^6$ direction
since $w_+$ and $w_-$ diverge as $\sim \mu$ in the limit.
The configuration with two infinite components
present for $m^{N_f}=(\lambda\Lambda^2)^{N_f}$
has a limit given by
\beq
C_L\left\{
\begin{array}{l}
t=1,\\
v=0,
\end{array}
\right.
\qquad
C_R\left\{
\begin{array}{l}
t=(\lambda\Lambda^2)^{-2N_f}(v^2+m^2)^{N_f},\\
w=0.
\end{array}
\right.
\label{iyitvac1}
\eeq
If we tune the parameter $m$ so that it approaches $\lambda\Lambda^2$
(or its $\Z_{N_f}$ phase rotation) as
$m=\lambda\Lambda^2-\sigma/\mu$, one of the $r=0$ solutions
is given by $w_+=i\mu\lambda\Lambda^2$, $w_-=-i\sigma$
and the $\mu\to\infty$
limit is a curve with a single component given by
\beq
\left\{
\begin{array}{l}
vw=\lambda\Lambda^2\sigma,\\[0.1cm]
t=(\lambda\Lambda^2)^{-2N_f}(v^2+m^2)^{N_f}.
\end{array}
\right.
\label{iyitvac2}
\eeq
We note that (\ref{iyitvac1}) can be obtained from (\ref{iyitvac2})
by taking the limit $\sigma\to 0$.
These supersymmetric configurations correspond to
the supersymmetric vacua $S=\sigma J$ of the IYIT model with
$m^{N_f}=(\lambda\Lambda^2)^{N_f}$.

\subsection*{\sl The Type IIA Limit}

\begin{figure}[htb]
\begin{center}
\epsfxsize=3.5in\leavevmode\epsfbox{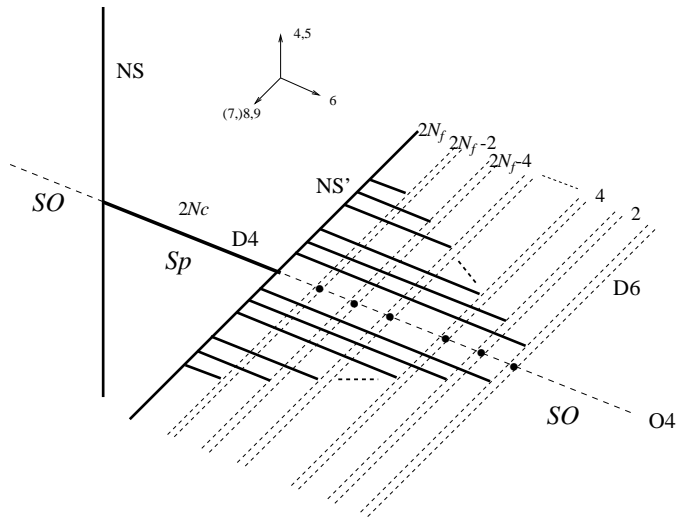}
\end{center}
\caption{Type IIA Configuration for IYIT Model}
\label{IYITIIA}
\end{figure}

We have proposed a realization of the (perturbed) IYIT model
on the worldvolume of the fivebrane in $M$ theory.
In the construction, we used a fictitious
duality between $Sp(N_c)$ and $Sp(-1)$ ``gauge'' theories as a guide
to read off the boundary condition.
Although we have observed some quantitative and qualitative
agreement with field theory about the
supersymmetric ground states,
it would be better to have more evidence for the proposal.
One obvious thing to check is whether we can see the elementary fields
of the gauge theory by considering the weak coupling Type IIA limit.
In Type IIA string theory,  the IYIT model (e.g. for $m=0$)
may be realized as the worldvolume theory of the brane configuration
as depicted in Figure \ref{IYITIIA}.
We can see the $Sp(N_c)$ gauge symmetry on the D4-branes stretched
between the NS and NS${}^{\prime}$ 5-branes, and the fundamental
chiral multiplets are created by open strings ending on these
D4-branes and the $2N_f$ D4-branes ending on the NS${}^{\prime}$ brane
from the right. The fluctuations of the singlet $S_{ij}$
correspond to the motion in the $x^{(7),8,9}$ directions of the
D4-branes on the right of the NS${}^{\prime}$ 5-brane. 
By taking a suitable Type IIA limit of the above
factorized configuration (which is a ${\it t}$-configuration for $m=0$),
we can actually see this kind of intersecting brane
configuration.
See \cite{H} for details.

\subsection{Non-supersymmetric Stable Vacuum}

\newcommand{\bj}{{\bar \jmath}}

Let us study the case with $m^{N_f}\ne (\lambda\Lambda^2)^{N_f}$
in detail. As we have seen above, in the $\mu\to\infty$ limit
all supersymmetric configurations go away from the region
with finite $vw$ and become infinitely elongated
in more than four directions.
We have already seen a similar phenomenon in the study of
fivebrane configurations corresponding to
$SU(N_c)$ SQCD with $N_f<N_c$ flavors
in which there is no stable vacuum 
\cite{HOO}.
However, there is a clear difference that distinguishes
the present case from such examples:
The holomorphic curve
(\ref{facto})
remains in the finite region and does not become infinitely
elongated, although it does not define a supersymmetric vacuum
because it contains a {\it t}-configuration.
We discuss what this fact implies for the issue of existence
of stable vacua.

As the classical Nambu-Goto action of the fivebrane shows,
the area of the real two-dimensional surface $\Sigma$
on which the fivebrane wraps
plays the role of the potential energy.
As far as the characteristic length scale of the brane
is much larger than the eleven-dimensional Planck length,
the classical supergravity is a good description and
an area-minimizing surface can be considered
as defining a stable vacuum as stated in section 2.
Therefore
one is tempted to look for an area-minimizing surface
which satisfies the boundary condition
given by (\ref{bc1})-(\ref{bc3}) or its $\mu\to\infty$ limit.
Of course, when the fivebrane approaches the $\Z_2$ fixed
plane one has to take the effects of the fixed plane into account.
The hypothesis that a \tc is not supersymmetric suggests
that there is an extra potential energy for a
{\it t}-configuration in addition to the energy associated with the
area of the surface. In this paper, we assume that 
this energy is negligible compared to the energy coming from the area
as long as the fivebrane is separated
from the $\Z_2$ fixed plane by a distance much larger than the
eleven-dimensional Planck length.

Although the surface we are considering is non-compact,
one can define a regularized area as far as the surfaces
obey a certain boundary condition at infinity.
A formal expression of the area of a surface $\Sigma$
embedded in the seven-dimensional part
of the space-time (transverse to the $\R^4$ direction
which the four-dimensional part of the fivebrane spans)
is given by
\beq
{\rm Area}(\Sigma)=\int_{\Sigma}\sqrt{g}\dd^2 x\,,
\eeq
where $\sqrt{g}\dd^2 x$ is the area element of the metric $g_{\mu\nu}$
on the surface induced from the seven-manifold.
In the present case, the seven manifold is the product of
the Taub-NUT space parametrized by $(t,v)$ and $\R^3$ parametrized by
$(w^7,w=w^8+iw^9)$.
Since we are interested in area-minimizing surfaces
under the condition that $w^7\to 0$ in each asymptotic region,
we may as well consider only the surfaces with $w^7\equiv 0$.
Namely, we consider $\Sigma$ to be a surface embedded in
the direct product of the Taub-NUT space and
the flat $w$-plane which is a manifold with
a K\"ahler metric $G_{i\bj}$
where
we use $i,j$ ($\bar \imath,\bj$) to denote the
indices of local
(anti-)holomorphic coordinates.
The induced metric $g_{\mu\nu}$ defines a complex structure
on $\Sigma$. Then, the area element can be expressed as
\beqa
\sqrt{g}\dd^2 x&=&
g_{z\bar z}\dd^2 z=
G_{i\bj}\left(\partial_z X^i\partial_{\bar z}X^{\bj}
+\partial_{\bar z}X^i\partial_{z}X^{\bj}\right)
\dd^2 z\nonumber\\
&=&2G_{i\bj}\,\partial_{\bar z}X^i\partial_{z}X^{\bj}\,\dd^2z
+{i\over 2}G_{i\bj}\dd X^i\wedge \dd X^{\bj}\,,
\label{areael}
\eeqa
where $z$ is a local complex coordinate of $\Sigma$.
The integral of the first term is finite if the boundary
condition at infinity is holomorphic and if the deviation
$\partial_{\bar z}X^i$ from the holomorphic embedding falls off
sufficiently fast.
To put it more appropriately, we include the finiteness
of the integral of this term as one of the boundary conditions.
The second term of (\ref{areael}) is the K\"ahler
form of the space-time restricted to the surface $\Sigma$.
Since the K\"ahler form is a closed two-form, 
its integral does not change
for a continuous variation of the surface and we may consider
it as a constant.
Thus, we may as well take
\beq
{\rm Area}^{\prime}(\Sigma)=2\int_{\Sigma}
G_{i\bj}\,\partial_{\bar z}X^i\partial_zX^{\bj}\,\dd^2z\,,
\label{areap}
\eeq
as the definition of the area as long as we are considering
surfaces in a given homology class.

Actually, the integral of the second term of (\ref{areael})
is divergent and needs some special care in order to
show that it can really be considered as a constant.
We first regularize the integral by cutting off some part of the
surface at infinity. In general, a variation of the surface
induces a change in the boundary of the cut-off surface,
and the integral might change by a boundary term.
The integral can be considered as a constant only when
the boundary term vanishes as we take the limit where
the cut-off is removed.
 For the configurations we are studying
there is a certain regularization such that
the boundary term does vanish as we remove the cut-off,
as we now show in the $\mu=\infty$ case
(the generalization to the case with finite $\mu$ is obvious).
In the asymptotic region (\ref{bc1}) where
the surface is parametrized by $w$,
we cut off the part with $|w|>W$ for some large $W$. Similarly
we cut off $|v|>V$ in the other asymptotic region (\ref{bc2})
for some large $V$.
It is easy to see that the boundary term vanishes in the
$V,W\to\infty$ limit.
 For example, let us consider the boundary term in the asymptotic region
(\ref{bc1}).
In the region with large $w$,
the surface is considered as a graph of the functions
$v(w),t(w)$, and the variation $\Sigma_0\to\Sigma_1$ of the
surface is represented by the
variation of such functions $v_0(w),t_0(w)\to v_1(w),t_1(w)$.
The boundary term is the integral of the K\"ahler
form on the boundary surface at $|w|=W$.
As the boundary surface, one can take for example
$w=W\e^{i\sigma}$,
$v=v_0(w)+\tau (v_1(w)-v_0(w))$,
$t=t_0(w)+\tau (t_1(w)-t_0(w))$
parametrized by $0\leq \sigma<2\pi$, $0\leq \tau\leq 1$.
Since the K\"ahler form is the sum of
$i\dd w\wedge\dd \bar w$ and
the K\"ahler form of the Taub-NUT space which is
well-parametrized by $v$ and $t$ near $t=1$ and $v=0$,
the boundary term vanishes in the limit $W\to\infty$
as long as $|v_1(w)-v_0(w)|,|t_1(w)-t_0(w)|\to 0$
which hold under the boundary conditions (\ref{bc1}).
Thus, one can really take (\ref{areap}) as the definition of the
regularized area.

The basic property of the regularized area (\ref{areap}) is that
it vanishes only for holomorphic curves.
Therefore, {\it if there is a holomorphic curve
satisfying the boundary condition,
an area minimizing surface $\Sigma$ in the same homology class
must be holomorphic} since
the holomorphic curve has ${\rm Area}^{\prime}=0$ and hence
an area minimizing surface must also have ${\rm Area}^{\prime}=0$
which implies that it is holomorphic.

\begin{figure}[htb]
\begin{center}
\epsfxsize=4.5in\leavevmode\epsfbox{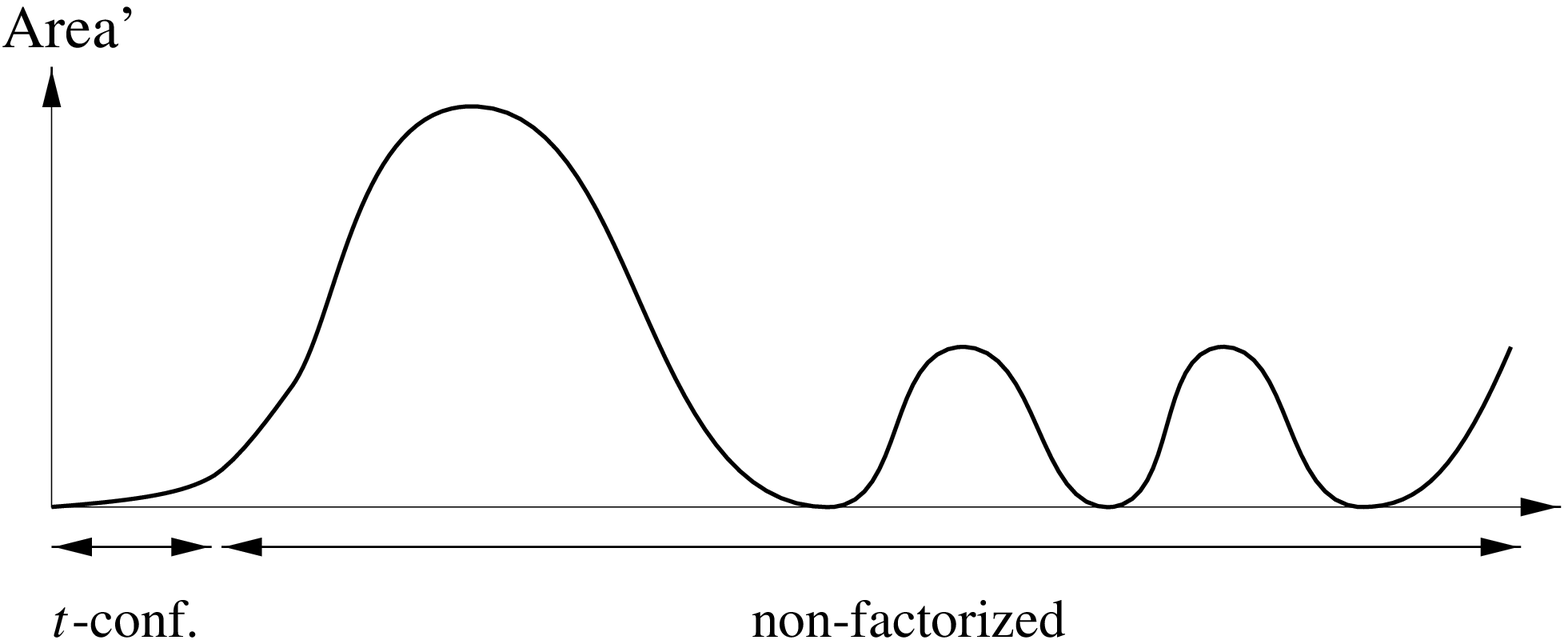}
\end{center}
\caption{The Graph of Area${}^{\prime}$ }
\label{finmu}
\end{figure}

 For finite $\mu$, there are various holomorphic curves satisfying
the boundary condition. All these
including the factorized one (\ref{facto})
are connected by a continuous deformation,
as can be seen as follows.
Since the space-time is topologically trivial in the
$w$ direction, we only have to show that their projections to the
Taub-NUT space are connected by a deformation.
The projection of the curve (\ref{vweq})-(\ref{tweq})
at the $r$-th branch, including the $\CP^1$
components, is described by the equation
\beq
x+y=(\lambda\Lambda^2)^{-2N_f}(v^2+m^2)^{N_f}+u_1(v^2+m^2)^{N_f-1}
+\cdots +u_{N_f-r}(v^2+m^2)^{r},
\eeq
for certain $u_i$'s,
while the factorized curve (\ref{facto}) projects to
\beq
x+y=(\lambda\Lambda^2)^{-2N_f}(v^2+m^2)^{N_f}+1.
\eeq
It is evident that these are connected
by a continuous deformation that does not change the
asymptotic behavior.
It is not clear whether this implies that we should
only take into account the surfaces in the same
homology class, but we shall assume this to be the case. 
Then, all the holomorphic
curves have the same minimum area
${\rm Area}^{\prime}=0$ and all other surfaces
have ${\rm Area}^{\prime}>0$.
The graph of the area functional is
thus schematically given by figure \ref{finmu}.

\noindent
The horizontal axis parametrizes the space of surfaces.
In the region called ``{\it t}-conf.'',
the surface contains a {\it t}-configuration
while the surface does not intersect
the $\Z_2$ fixed plane in the region called ``non-factorized''.
In the transition region between them,
two parts of the surface intersect
the fixed plane at the same point.
The minima correspond to the holomorphic curves.
The minima with non-factorized curves run
away to infinity in the $\mu\to\infty$
limit, but the minimum in the ``{\it t}-conf.'' region
is the factorized curve (\ref{facto}) and does not
go away.

Our basic hypothesis that a \tc is not supersymmetric
means that the configuration with factorized holomorphic curve
has positive energy compared to the non-factorized holomorphic curves,
although it minimizes the area.
More generally, such an extra energy should be positive in the
whole ``{\it t}-conf.'' region, fall down toward
the transition region, and be negligible in
the part of the ``non-factorized''
region in which the curve is separated from the $\Z_2$ fixed plane by
a distance much larger than the eleven-dimensional Planck scale.
The competition between the force from this extra energy
and the force from the gradient of ${\rm Area}^{\prime}$
determine the location of a possible stable configuration.

\begin{figure}[htb]
\begin{center}
\epsfxsize=2.0in\leavevmode\epsfbox{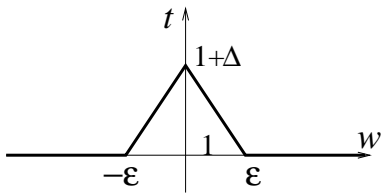}
\end{center}
\caption{A One-Dimensional Analog}
\label{oned}
\end{figure}

Therefore,
it is important to know how fast the area grows as
the factorized curve (\ref{facto}) is deformed so that
the two intersection points approach each other.
Namely, we want to find an area minimizing surface
with a given distance between the two components and compute the area
as a function of the distance.
 For simplicity let us consider the $\mu=\infty$ case.
The component $C_L$ is at $t=1,v=0$ and intersects the $\Z_2$ fixed
plane $v=w=0$ at $t=1$ while the component $C_R$ is at
$t=(\lambda\Lambda^2)^{-2N_f}(v^2+m^2)^{N_f}$, $w=0$ and intersects
the $\Z_2$ fixed plane at $t=(m/\lambda\Lambda^2)^{2N_f}$.
We want for example to deform $C_L$ by replacing $t\equiv 1$
by $t=t(w)$, a function of $w$, so that the area is minimized
under the condition $t(w)\to 1$ as $w\to\infty$ and $t(0)=1+\Delta$
for a fixed $\Delta$.
One can consider an analogous problem in the simplified
one-dimensional analog where the
parameters $t,w$ are considered as real numbers and
minimum area is replaced by minimum length.
In this one-dimensional analog,
for the configuration given by $t(w)\equiv 1$ for
$|w|\geq\epsilon$, and $t(w)=1+\Delta(1-|w|/\epsilon)$ for
$|w|\leq\epsilon$ (see figure \ref{oned}),
the length decreases
as $\epsilon$ is increased and does
not attain a minimum as the $\epsilon\to\infty$ limit no longer
satisfies the required asymptotic behavior.

One may wonder whether our two-dimensional
model exhibits this ``run-away'' behavior.
To examine this, we compute the area of the surface
in the Euclidean space $\C^2$ given by
the same function $t(w)$ as above where now we consider $t,w$ to be
complex variables. One may be worried by the fact that the surface
is not smooth at $|w|=0$ and $\epsilon$ because a singularity might
suggest an instability of the configuration,
but such a singularity can be smoothed out without change of the area.
The difference of the area
from the one of the flat surface is given by
\beq
{\sl \Delta}{\rm Area}=
\left(\sqrt{1+\Delta^2/\epsilon^2}-1\right)\pi
\epsilon^2\,.
\label{ardif}
\eeq
This is positive and is
monotonically {\it increasing} as a function of $\epsilon$.
Therefore, in the two-dimensional case,
we may not have to worry about such a run-away behavior for
$\epsilon\to\infty$
as in the case of one-dimensional analog.

However, there is another problem.
The difference ${\sl \Delta}{\rm Area}$ (\ref{ardif})
decreases as we decrease $\epsilon$ and
approaches zero as $\epsilon\to 0$ but the $\epsilon=0$ surface
has a spike singularity which cannot be smoothed out without 
increasing the area.
Namely, for any fixed $\Delta$,
the area can be made arbitrarily small 
but no smooth surface can attain ${\rm Area}^{\prime}=0$.
In particular, this is also true for the case
$\Delta=(m/\lambda\Lambda^2)^{2N_f}-1:=\Delta_*$
in which the two intersection points coincide.
However, we cannot conclude from this that there is no
local minimum of the energy. This is because a surface with sufficiently
small ${\rm Area}^{\prime}$ has characteristic length $\epsilon$
smaller than the eleven-dimensional Planck length,
and our argument (based on the low energy action)
does not apply to such surfaces.
Precisely because of this reason, we cannot make a decisive
statement concerning the stable vacuum within the eleven-dimensional
supergravity approximation of $M$ theory.

We have found that the slope of ${\rm Area}^{\prime}$ is almost zero
as a function of the distance between the two intersection points.
On the other hand, we expect a rapid growth of the extra energy
as the two intersection points are separated: Otherwise,
there would be an extra light mode in the examples we have
considered in the previous section which is absent in the
corresponding $Sp(N_c)$ gauge theories.
Therefore, it is unlikely that the potential minimum
is in the ``{\it t}-conf.'' region, although we need a more careful
estimate of the growth of the extra energy
in order to completely exclude this possibility.

\begin{figure}[htb]
\begin{center}
\epsfxsize=2.0in\leavevmode\epsfbox{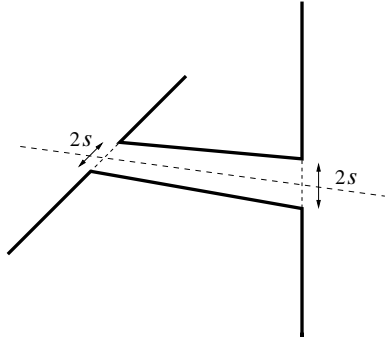}
\end{center}
\caption{A Deformation From The Spike Surface}
\label{rise}
\end{figure}

The fact that ${\sl \Delta}{\rm Area}$ (\ref{ardif}) approaches zero
when approaching the surface with a spike
singularity shows that ${\rm Area}^{\prime}$ in the ``non-factorized''
region also falls off when approaching the spike surface with
$\Delta=\Delta_*$.
In other words, ${\rm Area}^{\prime}$
grows when going away from the spike surface in the direction to the
``non-factorized'' region.
In order to illustrate this, we consider a family of configurations
parametrized by $s$ in the $\mu=\infty$ case
as depicted in figure \ref{rise}.\footnote{
We thank Michael Peskin for asking a question which lead to this
computation.}
As $s\to 0$, the configurations approach
the surface with the spike singularity.
To be more precise, we construct the configurations
from the factorized curve
(\ref{facto}) by cutting off the two discs --- $|w|\leq s$ of $C_L$
and $|v|\leq s$ of $C_R$ --- and connect the boundary circles
by a cylinder
(for simplicity of the discussion, we assume that $v$, $w$
and $t$ are coordinates of the flat Euclidean space $\C^3$
with the metric given by $|\dd v|^2+|\dd w|^2+|\dd t|^2$).
As the cylinder, we can choose
$t=1+\Delta_*\tau$, $v=sf(\tau)\e^{i\theta}$ and
$w=sg(\tau)\e^{-i\theta}$, where $0\leq\tau\leq 1$, $0\leq\theta\leq
2\pi$ are the coordinates of the cylinder,
and $f(\tau)$ (resp. $g(\tau)$)
is some non-negative function starting from $f(0)=0$ and ending at
$f(1)=1$ (resp. from $g(0)=1$ to $g(1)=0$) such that $f^2+g^2$ is
always positive.
Then the ${\rm Area}^{\prime}$ is given by
\beq
{\rm Area}^{\prime}
=2\pi s^2\int_0^1\sqrt{\Delta_*^2/s^2+(f^{\prime})^2
+(g^{\prime})^2}\sqrt{f^2+g^2}\dd\tau \,-\,2\pi s^2.
\eeq
One can show by using the inequality
$\sqrt{(f^{\prime})^2+(g^{\prime})^2}\sqrt{f^2+g^2}
\geq ff^{\prime}-gg^{\prime}$
that this is indeed non-negative.
Since it is zero at $s=0$, it can never decrease as $s$ is increased.
One can actually show
using the same inequality that it is
monotonically increasing as a function of $s$ at {\it all}
values of $s\geq 0$
for an arbitrary choice of the differentiable functions
$f(\tau)$, $g(\tau)$.
The parameter $s$ can be considered as a counterpart of the
eigenvalue $\sigma$ of the singlet $S_{ij}$,
as the supersymmetric configuration (\ref{iyitvac2})
in the $\Delta_*=0$ case suggests.
The monotonic growth of ${\rm Area}^{\prime}$ as $s\to$ large
may be considered as an analog of the potential growth as
$\sigma\to$ large in the field theory which can be shown
near the ultra-violet cut-off $\lambda \sigma\lsim\MUV$
by the one loop computation (and can be continued to smaller values of
$\sigma$ to some extent by a renormalization group argument).
In the brane picture the ${\rm Area}^{\prime}$
itself continues to fall off as $s\to$ small up until $s=0$.
Of course,
${\rm Area}^{\prime}$ can really be considered as the
potential energy only if the length scale set by $s$ is much larger
than the eleven-dimensional Planck length.

 For finite $\mu$, ${\rm Area}^{\prime}$
also decreases when approaching any of the non-factorized holomorphic
curves, but these go away to infinity in the $\mu\to\infty$ limit,
while the fall-off toward the spike surface
remains as we have illustrated above.
In addition, there are no other obvious directions in which the
potential decreases. 
This may suggest that, in the limit $\mu\to\infty$,
there is a minimum of the potential energy
in a region near the singular surface.
In field theory, it was difficult to obtain definite
information about the stable vacuum, such as its location or vacuum
energy,  because of the difficulty
in analyzing the K\"ahler potential of $\sigma$ in the region
where $\sigma$ is small.
In the brane picture, the difficulty is in dealing with
the fivebrane which has a characteristic length scale smaller than
the Planck length, for which we need
information about the
fivebrane dynamics in \MT
beyond the eleven-dimensional supergravity approximation.

\subsection*{Remark}

In \cite{KS}, the IYIT model was realized in $SO(32)$ heterotic string
theory compactified on a K3-fibred Calabi-Yau three-fold by encoding the 
$SU(2)$ gauge symmetry in a singularity of the gauge bundle
along a section (a Riemann surface) of the K3-fibration. It would be
interesting to find a relation with our construction of the model.
However, there is a clear difference from ours.
The parameter corresponding to the dynamical scale $\Lambda$
is related by $\sqrt{\alpha^{\prime}}\Lambda=\e^{-R^2/\alpha^{\prime}}$
to the size $R$ of the Riemann surface, which is one
of the moduli of string theory. Since the vacuum energy would be
$\Lambda^4$ times some dimensionless quantity, this may cause a
problem of vacuum instability.
In our case, it takes an infinite amount of energy to vary
the parameter $\Lambda$ because a change of $\Lambda$ causes a change of the
asymptotic boundary condition. Therefore, a map
between our model and that of \cite{KS}, if exists, would be realized in some
limit which freezes the degrees of freedom 
corresponding to the variation of the size of the Riemann surface
on the heterotic side.

\section{$Sp \times SO$ - Run away Behavior}

In this section we will study the IYIT model with a gauged flavor group.
This theory has several features in common with the original IYIT model,
but there is one crucial difference. In the case where $N_f=N_c+1$, 
supersymmetry is still broken, but there no longer exists a stable
non-supersymmetric vacuum. It is therefore very interesting to compare 
the brane geometry for the gauged model to the one for the original IYIT
model discussed in the previous section. It may provide us with a general
rule how to distinguish between runaway behavior and the existence
of stable non-supersymmetric vacua, given some brane configuration for
a theory with dynamical supersymmetry braking. 

\subsection{Field Theory}

One way to construct a brane configuration for an $N=1$ theory is to start
with an $N=2$ theory and to introduce suitable mass terms so that when
we send the masses to infinity we recover the $N=1$ theory. We used this
in the previous sections to study $N=2$ QCD with gauge group $Sp(N_c)$
as well as the IYIT model. The advantage
of this method is that one can see which $N=1$ theories have a supersymmetric
vacuum and which ones do not. If the brane configuration has a well-defined
limit when we send the masses to infinity, then this is the brane configuration
for the corresponding $N=1$ theory. If no such well-defined limit exists, supersymmetry
is generically broken in the $N=1$ theory. Clearly, this argument is only valid
if supersymmetry is unbroken for finite values of the masses. To facilitate 
comparison with the brane geometry, we will therefore discuss an $N=2$
version of the gauged IYIT model, broken to $N=1$ by mass terms.

We thus consider an $N=2$ gauge theory with gauge group $Sp(N_c) \times SO(2N_f)$,
and one hypermultiplet $Q_{ai}$ transforming in the `bifundamental' representation 
$({\bf 2N_c},{\bf 2N_f})$. From the $N=1$ point of view there are two chiral
superfields $(\Phi_{\rm Sp})_{ab},
(\Phi_{\rm SO})_{ij}$ transforming in the adjoint representations of $Sp(N_c)$ and 
$SO(2N_f)$ respectively. The superpotential, including a mass term for
$\Phi_{\rm Sp}$,  reads
\be \label{pot}
W =  \Tr(Q^t J \Phi_{\rm Sp}J Q) + \Tr(J Q \Phi_{\rm SO}
Q^t) + \mu \Tr(J\Phi_{\rm Sp}J \Phi_{\rm Sp}).
\ee
Under gauge transformations $Q \rightarrow gQh^t$, $\Phi_{\rm SO} 
\rightarrow h \Phi_{\rm SO} h^t$ and $\Phi_{\rm Sp} \rightarrow 
g \Phi_{\rm Sp} g^t$, where $g\in Sp(N_c)$ and $h\in SO(2N_f)$. 
 For large $\mu$ we can integrate out $\Phi_{\rm Sp}$. Introducing
the meson field $M=Q^tJQ$, the resulting superpotential after integration
reads
\be \label{pot2}
W = -\frac{1}{4\mu} \Tr(MM) + \Tr(M \Phi_{\rm SO}).
\ee
To analyze what happens in the quantum theory it is convenient to
know the charges of the various fields and parameters under the 
global $U(1)_A \times U(1)_R$ symmetry. They are given by
\be
\begin{array}{cccccc}
 & Q & \Lambda_{N=1,Sp}^{3(N_c+1)-N_f} & \mu & \Phi_{\rm SO} &
 \Lambda_{N=2,SO} \\
U(1)_A & 1 & 2N_f & 4 & -2 & -2 \\
U(1)_R & 1-\frac{N_c+1}{N_f} & 0 & 2-4\frac{N_c+1}{N_f} & 
2 \frac{N_c+1}{N_f} & 2 \frac{N_c+1}{N_f}. 
\end{array}
\ee
The $U(1)_R$ charges have been chosen so that they yield the usual  
$U(1)_R$ charge assignments for the $Sp(N_c)$ theory.

We now discuss the four different cases $N_f\leq N_c$, $N_f=N_c+1$,
$N_f=N_c+2$ and $N_f>N_c+2$.  

\underline{$N_f \leq N_c$}. The classical moduli space is
given by $M=0$ while $\Phi_{\rm SO}$ takes values in the
Cartan subalgebra. In other words, it is just the moduli
space of the $N=2$ $SO(2n_f)$ theory. In the quantum theory, strong coupling
dynamics can generate corrections to the superpotential. The global
symmetry restricts the form of such a correction to a flavor invariant
combination of the form 
\be \label{wdy0}
W\sim \mu^{-\alpha}
\left(\frac{\Lambda_{N=1,Sp}^{3(N_c+1)-N_f}}{ M^{N_f}}
\right)^{\beta/(N_c+1-N_f)}
\Lambda_{N=2,SO}^{\gamma} M^{1-\beta+\alpha} \Phi_{\rm
SO}^{1-\alpha-\beta-\gamma}.
\ee
If all the nonperturbative dynamics is due to instanton effects then
$\beta$ and $\gamma$ must be nonnegative integers. Because of
holomorphy, $\alpha$ must also be a nonnegative integer. The dynamics
of the $SO(2N_f)$ gauge group does not generate any 
non-perturbative superpotential because from that point of view
the theory looks like an $N=2$ theory. The dynamics of the
$SP(N_c)$ theory is that of an $N=1$ theory with massive quarks.
In such a theory a non-perturbative superpotential is generated, which
is given by  \cite{IP},
\be \label{wdy}
W_{dyn} \sim \left(\frac{\Lambda_{N=1,Sp}^{3(N_c+1)-N_f}}{{\rm Pf}
 \,\, M} \right)^{1/(N_c+1-N_f)}.
\ee
The full superpotential $W$ is the sum of (\ref{pot2}) and (\ref{wdy}).

An alternative derivation of the superpotential is obtained by
putting a coefficient $\lambda$ in front of
the second term in (\ref{pot2}). For $\lambda\rightarrow 0$ the
theory is well-behaved and the full superpotential should therefore
be analytic in $\lambda$. This imposes the restriction
$\alpha+\beta+\gamma \leq 1$ in (\ref{wdy0}). The term with
$\gamma=1$, $\alpha=\beta=0$ cannot be present as it cannot be
made invariant under $SO(2N_f)$. The remaining possibilities 
correspond to the three terms in $W=$(\ref{pot2})$+$(\ref{wdy}).

The full superpotential implies that supersymmetry is dynamically
broken, as the $\partial W/\partial \Phi_{\rm SO}=M=0$ equation 
is incompatible with the equation $\partial W/\partial M=0$ due to
(\ref{wdy}). We will later see a confirmation of this in the brane
analysis. A similar situation appears in the case of $SU$ gauge groups, see
\cite{gipe}. 

\underline{$N_f= N_c+1$}. For this value of $N_f$, global symmetries
restrict a contribution to the superpotential to be of the form
\be
\mu^{-\alpha} \Lambda_{N=1,Sp}^{2\beta} \Lambda_{N=2,SO}^{\gamma}
M^{1+\alpha-\beta} \Phi_{\rm SO}^{1-\alpha-\gamma}.
\ee
By going to large $\Phi_{\rm SO}$ we can see that none of the possible
terms will be generated, as the only dynamics is that of a massive 
$N=1$ $SP(N_c)$, plus that of some $U(1)$ $N=2$ vector multiplets.
The only nontrivial dynamics comes from the $Sp(N_c)$ gauge
group, which generates the  quantum
constraint ${\rm Pf}\,M =\Lambda_{N=1,Sp}^{2N_f}$. This is incompatible
with the equation $M=0$ obtained from the tree level superpotential 
(\ref{pot2}), and supersymmetry is dynamically broken. To analyze
whether there is a stable vacuum or not, we focus in analogy to
section four on the region where $\Phi_{SO}$ is large. In that region
the quarks acquire a large mass and can therefore be integrated out.
What remains is a pure $Sp(N_c)$ gauge theory with a scale
$\Lambda_{N=1}$, together with an
$SO(2N_f)$ gauge theory with the adjoint matter field $\Phi_{SO}$. 
Gaugino condensation in the pure $Sp(N_c)$ gauge theory generates
a superpotential $W\sim \Lambda_{N=1}^3$. Using the scale matching
relation $\Lambda_{N=1}^{3(N_c+1)}={\rm Pf}(\Phi_{\rm SO})
\Lambda_{N=1,Sp}^{3(N_c+1)-N_f}$ we can write this as
\be \label{spot2}
W\sim \left( {\rm Pf}(\Phi_{\rm SO}) \right)^{1/N_f}
\Lambda_{N=1,Sp}^2.
\ee
The scalar potential is given by $V\sim \left| \frac{\partial W}{\partial
\Phi_{\rm SO} } \right|^2 g^{\Phi_{\rm SO} \bar{\Phi}_{\rm SO}}$,
where $ g^{\Phi_{\rm SO} \bar{\Phi}_{\rm SO}}$ is the inverse
of the metric appearing in the kinetic term for $\Phi_{\rm SO}$. 
Since $\partial W/\partial \Phi_{\rm SO}$ is roughly of order one,
the behavior of the scalar potential for large $\Phi_{\rm SO}$ is
determined by the behavior of 
$ g_{\Phi_{\rm SO} \bar{\Phi}_{\rm SO}}$ for large $\Phi_{\rm SO}$..
This can be estimated using a one-loop calculation. The result of the 
one-loop calculation reads
\be
 g_{\Phi_{\rm SO} \bar{\Phi}_{\rm SO}} \sim 1 + \frac{1}{8\pi^2}
(2(2N_f-2)-2N_c) \log
\left| \frac{\Phi_{\rm SO}}{M_{UV}} \right|
\ee
with $M_{UV}$ the ultra-violet cut off. For large $|\Phi_{\rm SO}|$ we see that
the metric becomes very large, and therefore the scalar potential has a 
runaway behavior and approaches
zero at infinity (a similar behavior also occurs for
$N_f<N_c+1$). This fact shows that the true minimum of the scalar potential
is at infinity, and that there is no stable vacuum for large 
$|\Phi_{\rm SO}|$. This argument does not exclude the possibility of
a local minimum in the scalar potential, but so far no evidence for
the existence of such local minima has been found. The brane analysis 
will in fact provide evidence to the contrary.

\underline{$N_f=N_c+2$}. The full superpotential is given by 
$W=$(\ref{pot2})$+$(\ref{wdy}). In the quantum theory $M$ is
an unconstrained field that can take on any value \cite{IP}.
The tree-level superpotential enforces $M=0$. Now, this point
is compatible with (\ref{wdy}) and part of the moduli space, it is
the special point for the $SP(N_c)$ theory where there is
confinement without chiral symmetry breaking. From the point
of view of the $SO$ gauge theory, both $\Phi$ and $M$ are massive,
because the tree-level superpotential equals
$-\frac{1}{4\mu}\Tr(M-2\mu\Phi_{SO})^2 + \mu\Tr\, \Phi_{SO}^2$.
Integrating them out leaves us with a pure $SO(2N_f)$ gauge
theory which has $2N_f-2$ discrete vacua. Thus, the total
number of discrete vacua is also $2N_f-2=2(N_c+1)$.

\underline{$N_f>N_c+2$}. At low energies the dynamics in the $Sp$ gauge
group is described by the dual magnetic theory with gauge group
$Sp(N_f-N_c-2)$ \cite{seib}. Using the dual description the tree-level
superpotential reads
\be
W_{tr}=\frac{1}{4\mu}\Tr(MM) + \frac{1}{\lambda} \Tr(JqMq^t) + 
\Tr(M\Phi_{SO}),
\ee
where $q$ are the dual magnetic quarks and $\lambda$ is an extra scale
that appears for dimensional reasons. The $SP(N_f-N_c-2)$ gauge group
generates a nonperturbative superpotential $W_{dy}$ which is again given by
(\ref{wdy}). The tree-level superpotential forces the 
theory in the non-abelian Coulomb point
$M=0$. At this point, the $SO$ part of the theory is
an $N=1$ $SO(2N_f)$ gauge theory with $2(N_f-N_c-2)$ massless
flavors, because $\Phi_{\rm SO}$ and $M$ are massive, but the 
magnetic quarks $q$ are massless.  This $SO(2N_f)$ theory
generates a non-perturbative superpotential due to gaugino
condensation \cite{inse}, and we see that supersymmetry is dynamically
broken.

\subsection{Branes}

To construct a brane configuration for the gauge theory with 
gauge group $Sp(N_c) \times SO(2N_f)$ and superpotential (\ref{pot}),
we first construct the brane configuration for the $N=2$ gauge theory
with $\mu=0$ and then try to introduce a mass term as in \cite{HOO}
by rotating one of the branes. The type IIA and M-theory brane 
construction for the $N=2$ gauge theory were discussed in \cite{lll}.
The type IIA construction consists of 3 NS 5-branes, with $2N_c$ D4
branes stretched between the first and second fivebrane, and $2N_f$
D4 branes stretched between the second and the third. In addition,
there is an O4 orientifold plane parallel to the D4 branes. With a
suitable choice of sign for the orientifold projection this describes
precisely the $N=2$ gauge theory of interest. After lifting the
description to M-theory, the corresponding fivebrane configuration is 
given by the equation
\be \label{fivebrane}
 F(t,v) \equiv \Lambda_{N=2,Sp}^{2(2N_c+2-N_f)} t^3 -
(v^2 P_{N_c}(v^2) + c)t^2 + Q_{N_f}(v^2) t - v^2 \Lambda_{N=2,SO}^{2(2N_f-2-N_c)}
=0,\ee
where $c$ is a constant (which can be determined by requiring that
(\ref{fivebrane})
is not a $t$-configuration) given by  
\be \label{c} 
c^2=4  \Lambda_{N=2,Sp}^{2(2N_c+2-N_f)} Q(0)
\ee
and $P_{N_c}$ and $Q_{N_f}$ are polynomials of order $N_c$ and $N_f$
in $v^2$. In the limit where $\Lambda_{N=2,SO}\rightarrow 0$, the
curve becomes a double cover of the curve for $Sp(N_c)$ gauge groups
with matter given in \cite{AS}, and similarly for 
$\Lambda_{N=2,Sp} \rightarrow 0$ we recover the curve for $SO(2N_f)$.

Recall that $v=x^4 + i x^5$ and that rotation of a curve means that
we rotate one of the NS fivebranes in the $w=x^8+i x^9$ direction. The
curve (\ref{fivebrane}) has three asymptotic regions where
$v\rightarrow \infty$ and $t\sim v^{2N_c+2}$, $t\sim v^{2N_f-2N_c-2}$
and $t \sim v^{2-2N_f}$ respectively, corresponding to the three
NS fivebranes. To give a mass to the adjoint $Sp(N_c)$ superfield
as in (\ref{pot}), we have to rotate the leftmost NS fivebrane (i.e.
the one where $t\sim v^{2N_c+2}$).
Rotating the middle NS fivebrane corresponds to giving masses to 
both $\Phi_{\rm Sp}$ and $\Phi_{\rm SO}$, and rotating the rightmost
NS fivebrane corresponds to giving a mass to $\Phi_{\rm SO}$ only. 

Because $w$ and $\mu$ are the only objects that transform nontrivial
under $U(1)$ rotations in the $8,9$ plane, we know that the projection
of the rotated curve in the $t,v$ plane must be given by the original
curve itself. Therefore, the rotated curve is given by the two
equations $F(t,v)=0$ and $w=w(t,v)$, and the goal is to determine
$w(t,v)$. If we compactify the original curve then $w$ would seem to 
be a meromorphic function on this compactified curve with a double
pole the location of the leftmost NS fivebrane. However, we have to
be more careful, because the space-time geometry is non-trivial due
to the presence of the O4 plane. The ${\bf Z}_2$ action induced by
the O4 plane maps $v\rightarrow -v$ and $w\rightarrow -w$. Modding
out by this ${\bf Z}_2$ creates an $A_1$ singularity in the $v,w$
space, and $v,w$ are not good coordinates on the quotient. Therefore,
we introduce the ${\bf Z}_2$ invariant coordinates
\be p=v^2,\qquad q=w^2,\qquad r=vw \ee
subject to the relation $pq=r^2$. The original curve is given by
an equation $F(p,t)=0$ and we are really looking for is functions
$q(p,t)$ and $r(p,t)$ subject to the relation $pq(p,t)  = r(p,t)^2$. 
In addition, if $p\rightarrow \infty$ and $t\sim p^{N_c+1}$, then
$r(p,t) \sim \mu p$, $q(p,t) \sim \mu^2 p$,
whereas in the other two asymptotic regions
as $p\rightarrow \infty$ we require $q(p,t)$ to remain finite. 
Thus, we see that $q(p,t)$ is a function on the compactified curve
with a single pole. The existence of such a function requires that
the compactified curve is a $\CP^1$. More precisely, there are
two possibilities. Either the compactified curve is a $\CP^1$,
or the compactified curve consists of several components, and
the component containing the point corresponding to 
the leftmost NS fivebrane is a $\CP^1$. We call these
two possibilities the non-factorized and factorized cases
respectively, and discuss them separately.

\underline{non-factorized curve}. In this case, the entire curve
must be a $\CP^1$, and the fact that $q(p,t)$ is a function with a single
pole on this $\CP^1$ implies that $q$ is a good global coordinate
on the entire curve. We should therefore be able to express $p$ and
$t$ in terms of $q$. At three points on the $\CP^1$ $p$ becomes
infinite and is a good local coordinate there; one of these
three points is $q=\infty$, the other correspond to finite values of 
$q$, and $p$ is therefore given by
\be p=\frac{(q-q_1)(q-q_2)(q-q_3)}{\mu^2(q-q_4)(q-q_5)}.
\ee
Because $\{q_1,q_2,q_3\} \cap \{q_4,q_5\} = \emptyset$,
$r(q)=\sqrt{p(q)q}$ always has a branch cut and is never globally
well-defined. We conclude that the rotation is impossible.

\underline{factorized curve}. It is straightforward to see that
$F(t,v)$ can be factored in terms of two polynomials in at most two
ways. Taking $\Lambda_{N=2,Sp}=\Lambda_{N=2,SO}=1$ for simplicity
these are
\bea \label{cas12}
1) & \quad & F(t,v)=(t+d)(t^2 - (v^2 P_{N_c}(v^2) + c +d) t -d^{-1}
v^2), \non \\
2) & \quad & F(t,v)=(t+dv^2) (t^2 -(v^2 P_{N_c}(v^2) + c + dv^2)t -
d^{-1}).
\eea
The first case requires $Q_{N_f}(v^2)=-d^{-1}v^2 -d (v^2 P_{N_c}(v^2)
+ c +d)$, so that $N_f=N_c+1$. In addition, according to (\ref{c}),
$c^2=4Q(0)=-4d(c+d)$, so that $c=-2d$. The first factor in $F(t,v)$
corresponds to the middle NS fivebrane, which we do not want to 
change\footnote{From the type IIA point of view, this factorization
corresponds to sticking 2 of the $2N_f$ D4 branes in the O4 plane so that
the charge of the O4 plane to the left and right of the middle NS
fivebrane is equal. The remaining D4 branes to the left and right
of the middle NS fivebrane can then be connected and subsequently one
can move the middle NS fivebrane away from the configuration.}. 
The second factor in $F(t,v)$ can not be further factorized and 
must therefore be a $\CP^1$. Applying the same logic as in 
the non-factorized case to this piece of the curve 
we find that $p=v^2$ must be of
the form $(q-q_1)(q-q_2)/\mu^2 (q-q_3)$. In order for 
$r=\sqrt{p(q)q}$ to be well-defined we need $q_1=q_2$ and $q_3=0$.
But this implies that there is only one value of $q$ for which $p=0$,
which contradicts the fact that there are two distinct values of
$t$ for $p=0$, namely $t=0$ and $t=-d$ ($d=0$ is clearly impossible).
Hence, the curve cannot be rotated.

It remains to analyze the second possibility in (\ref{cas12}). That one
requires $Q_{N_f}(v^2)=-d^{-1} -dv^2  (v^2 P_{N_c}(v^2) + c + dv^2)$,
fixing $N_f=N_c+2$. Furthermore, $c^2=4Q(0)=-4d^{-1}$. Again, the
first factor in (\ref{cas12}) corresponds to the middle NS fivebrane,
so that we have to require that the second factor corresponds to
a $\CP^1$. The second factor describes the curve of a pure
$Sp(N_c)$ gauge theory, and we already know from section~3 when this
curve can be rotated. There are precisely $N_c+1$ discrete
possibilities for the parameters in $P_{N_c}(v^2)$, corresponding to
the $N_c+1$ discrete vacua of pure $Sp(N_c)$ gauge theory. In addition
there are two choices for $c$, so that the total number of vacua that
survives the mass perturbation is $2(N_c+1)$. 

Comparing the results of the field theory with those of the brane
analysis, we see that there is a complete agreement.

\subsection{Runaway Behavior}

We have seen that the structure of the spaces of supersymmetric vacua
obtained from the brane and from field theory agree. It is an
interesting question whether the runaway behavior of the scalar 
potential that we discussed for $N_f=N_c+1$ can be seen in the brane
language as well. A priori this is a more difficult question.
In field theory, it involves knowledge of the K\"ahler potential.
As was demonstrated in \cite{kah}, it is very difficult to obtain
quantitative information about the K\"ahler potential using the
M-theory fivebrane in the eleven dimensional
supergravity limit of M theory. In the present case we do not need to know
the detailed quantitative structure of the K\"ahler potential,
we are only interested in some qualitative features, and one may
hope that the brane will still reproduce these qualitative features.

Let us first try to rephrase runaway behavior in the brane language.
To find a brane configuration for a given theory we first need to 
specify some boundary conditions. In the case we have been
considering              
so far, namely an $N=2$ gauge theory with gauge group $Sp(N_c) \times
SO(2N_f)$ broken to $N=1$ by a mass term for $\Phi_{\rm Sp}$, these boundary
conditions are that as $v\rightarrow \infty$, there are three distinct
asymptotic regions: (i)
 $t\sim v^{2N_c+2},w\sim \mu v$, (ii) $t\sim v^{2N_f-2N_c-2},
 w\rightarrow 0$
and (iii) $t \sim v^{2-2N_f}, w\rightarrow 0$. In addition there
are certain conditions that have to do with the M-theory O5 plane. As we
discussed previously, there should be no $t$-configuration, i.e.
there should be no transversal intersection
of a single fivebrane with the orientifold plane, the orientifold
plane should have the right fivebrane charge and the brane
configuration should be ${\bf Z}_2$ invariant. We will ignore the
subtleties associated to the orientifold plane for the time being and
first consider the case without an orientifold plane. Consider a given brane
configuration that satisfies the right boundary conditions and which has 
four trivial directions corresponding to the four dimensions where the
field theory lives. Such a brane configuration corresponds to a
particular field configuration in the field theory. If the brane 
corresponds to a supersymmetric cycle, it describes a supersymmetric
field configuration. From the fivebrane action it is clear that the
area of the fivebrane corresponds to the scalar potential of the field
theory, and a minimal area fivebrane configuration corresponds to
a minimum of the scalar potential. Therefore runaway
behavior should manifest itself in the brane picture in the following
way: given a brane
configuration that satisfies the right boundary conditions one should
always be able to deform it so as to reduce its area, without ever 
reaching a limiting configuration. If one keeps on reducing its area
some of the parameters describing the brane configuration will
runaway to infinity.

Let us now try to apply this picture to the case with $N_f=N_c+1$,
where we expect such runaway behavior from field theory. 

We first consider some general features of minimal surfaces in K\"ahler
manifolds with metric $G_{i\bj}$, where $X^i$ are the holomorphic
coordinates in target space. A surface is locally given by some
functions $X^i(z,\bar{z})$ depending on the complex coordinates
$z,\bar{z}$. The area of a surface is reparametrization invariant and we
will fix this reparametrization invariance by the Virasoro conditions
\be \label{vir} G_{i\bj} \partial X^i \partial X^{\bj} = 
G_{i\bj}\bar{\partial} X^i \bar{\partial} X^{\bj} = 0.
\ee
This has the advantage, familiar from string theory, that the area of the
surface is given by the simple expression
\be
S=\int_{\Sigma} d^2 z\, G_{i\bj} (\partial X^i \bar{\partial} X^{\bj} 
+ \bar{\partial} X^i \partial X^{\bj}).
\ee
As shown in section~4.3, this can be rewritten as
\be \label{form2}
S=2\int_{\Sigma} d^2 z\, G_{i\bj}  \bar{\partial} X^i \partial X^{\bj}
+ \int_{\Sigma} X^{\ast} \omega
\ee
where $\omega$ is the K\"ahler form of target space. We also showed that in 
order to find a minimal surface in a given homology class and with certain 
given boundary conditions,
we can consider the second term in (\ref{form2}) to be constant and drop it. 
We are thus left with the task of minimizing the regularized volume
\be \label{form3}
S={\rm Area}'(\Sigma)=
2\int_{\Sigma} d^2 z\, G_{i\bj}  \bar{\partial} X^i \partial X^{\bj}.
\ee
As we explained, an
advantage of this expression is that in the case where we are dealing with
holomorphic boundary conditions this expression is finite
(which is in some sense the definition of  holomorphic boundary conditions).
It also clearly shows that any holomorphic surface is automatically minimal.
The variation of $X^i$ in (\ref{form3}) leads to the minimal surface 
equation
\be \label{cond}
G_{i\bj} \partial \bar{\partial} X^{\bj} + 
\bar{\partial} X^{\bar{k}} \partial X^{\bj} \partial_{\bar{k}} 
G_{i\bj} = 0.
\ee

Let us now discuss two specific situations. First
consider the  case where $G_{i\bj}$ is constant. 
The minimal surface equation reads $\partial \bar{\partial} X^i = 
\partial\bar{\partial} X^{\bar{\imath}}=0$. This implies that 
$X^i=X^i(z) + X^i(\bar{z})$, where $X^i(z)$ is a possibly
multi-valued holomorphic function. Because (\ref{form3})
is finite, $X^i(\bar{z})$ must be bounded as one
approaches infinity. There are no bounded non-trivial
anti-holomorphic functions on a Riemann surface, even if
one allows for multi-valuedness. Therefore $X^i(\bar{z})$ 
must be identically zero and the minimal surface must 
necessarily be holomorphic. An alternative way to see this
is to partially integrate (\ref{form3}) and then to use the
minimal surface condition to show it vanishes. Because (\ref{form3}) is
positive definite and vanishes only for holomorphic curves,
the minimal surface must
be holomorphic. 

The second situation we want to consider is what happens when we have 
a holomorphic surface with given boundary conditions and then we change
the holomorphic boundary conditions, like turning on a mass $\mu$
for $\Phi_{\rm Sp}$. At infinity, the change in the boundary
conditions is very large. However, this is a change of holomorphic
boundary conditions only, so that we expect a large change in
the holomorphic part of $X^i$ and the anti-holomorphic part of
$X^{\bj}$, but only a very small change in the
anti-holomorphic part of $X^i$ and the holomorphic part of
$X^{\bj}$. Since (\ref{form3}) only involves the latter,
we can construct a minimal surface for small $\mu$ as
a small perturbation of the one for $\mu=0$. Indeed, if we substitute
$X^i\rightarrow X^i+\xi^i$, $X^{\bj} \rightarrow X^{\bj} + 
\xi^{\bj}$ in (\ref{cond}) and expand to first order in $\xi$
using the fact that $X^i$ is holomorphic and $X^{\bj}$ is
anti-holomorphic we get
\be \label{cond2}
G_{i\bj} \partial \bar{\partial} \xi^{\bj} + 
\bar{\partial} X^{\bar{k}} \partial \xi^{\bj} 
\partial_{\bar{k}} G_{i\bj} =0
\ee
which is only an equation for the very small quantity $\partial \xi^{\bj}$.
We can rewrite (\ref{cond2}) as $\bar{\partial} (G_{i\bj} \partial
\xi^{\bj} )=0$ which implies that
\be
\int d^2 z \, G_{i\bj} \bar{\partial} \xi^i \partial \xi^{\bj}
=0
\ee
and therefore $\xi^i$ has to be holomorphic. We conclude that for
sufficiently small $\mu$ any minimal surface must necessarily be
holomorphic. 

This statement could in principle get modified in the presence
of an orientifold, due to the additional potential energy it
creates. However, as long as there are no holomorpic $t$-configurations
after the perturbation, 
we believe we can restrict our attention to 
non-$t$-configurations only. The discussion above then still applies
in the presence of an orientifold, as long as the distance between
the fivebrane and the $\Z_2$ fixed plane is much larger than the
eleven-dimensional Planck length. In particular, in the presence
of an orientifold we cannot exclude the existence of a local
nonzero minimum of the scalar potential in the region where the
eleven dimensional supergravity approximation breaks down.
This is similar to the situation in field theory, where one
cannot say anything about the behavior of
the scalar potential at strong coupling
due to lack of knowledge of the K\"ahler potential in that region. 

We therefore see that for $N_f=N_c+1$ that 
for small enough $\mu$
(i) there is no holomorphic
surface satisfying the right boundary conditions and (ii) that any 
minimal surface has to be holomorphic. The only way to match these
two observations is if (\ref{form3}) can be made arbitrarily small, but
cannot be made equal to zero. A simple analogy of this situation is
the problem 
to find the shortest real curve in the $x,y$-plane that approaches $y=a$
as $x\rightarrow -\infty$, and approaches $y=b$ as 
$x\rightarrow +\infty$. Any curve that satisfies the right
boundary conditions is clearly not the straight line. We can
always make the curve straighter and straighter, thereby reducing
its length, but we never reach the point where it becomes a
straight line.

Exactly the same situation appears in the case at hand. Given some surface,
we can always reduce its area, for instance by replacing the surface inside
a large sphere by the minimal surface inside the sphere, keeping the
boundary conditions on the sphere fixed. We can make the sphere
arbitrarily large, but never reach an exact minimal surface.

It is now clear that this behavior is precisely the manifestation of
runaway behavior in the brane picture. It also shows that under
quite general circumstances (e.g. in the case of the product group for
all $N_f \neq N_c+2$) where supersymmetry is dynamically broken,
we do expect runaway behavior in the scalar potential. 
Our discussion of the case where $G_{i\bj}$ is constant
implies that similar things happen in the case of products
of $SU$ gauge groups. Thus, runaway behavior in the
presence of holomorphic boundary conditions seems to be a quite 
general phenomenon, except in situations with additional
interactions as discussed in section~4.
 Finding counterexamples where a non-holomorphic minimal surface 
satisfies holomorphic boundary conditions is obviously an
important problem, for which something quite non-trivial in
the brane picture must happen.

\noindent
{\bf Acknowledgement}

We thank Hitoshi Murayama for many useful discussions.
We furthermore would like to acknowledge discussions with
Amit Giveon, David Kutasov, 
Joe Lykken, Al Shapere and Scott Thomas.
We thank Savas Dimopoulos and Michael Peskin for remarks and
questions that lead to an improvement of the paper.

This work is supported in part by NSF grant PHY-951497 and DOE
grant DE-AC03-76SF00098.
JdB is a fellow of the Miller Institute for Basic Research
in Science.

\end{document}